# A Survey on Performance, Current and Future Usage of Vehicle-To-Everything Communication Standards

Falk Dettinger*;* Matthias Weiß; Daniel Dittler; Johannes Stümpfle; Maurice Artelt; Michael Weyrich

*Abstract—* Wireless communication between road users is essential for environmental perception, reasoning, and mission planning to enable fully autonomous vehicles, and thus improve road safety and transport efficiency. To enable collaborative driving, the concept of vehicle-to-Everything (V2X) has long been introduced to the industry. Within the last two decades, several communication standards have been developed based on IEEE 802.11p and cellular standards, namely Dedicated Short-Range Communication (DSRC), Intelligent Transportation System G5 (ITS-G5), and Cellular- and New Radio- Vehicle-to-Everything (C-V2X and NR-V2X). However, while there exists a high quantity of available publications concerning V2X and the analysis of the different standards, only few surveys exist that summarize these results. Furthermore, to our knowledge, no survey that provides an analysis about possible future trends and challenges for the global implementation of V2Xexists. Thus, this contribution provides a detailed survey on Vehicle-to-Everything communication standards, their performance, current and future applications, and associated challenges. Based on our research, we have identified several research gaps and provide a picture about the possible future of the Vehicle-to-Everything communication domain.

*Index Terms* — C-V2X, Challenges on V2X, DSRC, IEEE 802.11, ITS-G5, NR-V2X, Performance comparison, Vehicle-to-Everything

## I. INTRODUCTION

Autonomous vehicles have gained significant prominence in recent years, becoming a central focus of research in the fields of transportation and technology. A high autonomy level can help to increase on-road safety and transportation efficiency and thus reduce accidents and traffic jams, while optimizing the fuel economy and the travel time simultaneously [1]. To reach these goals, advanced vehicle functions with additional requirements regarding environmental perception and decision-making show up. This leads to increasing information requirements of the functions, necessitating a powerful exchange of data and information. Overall, the information exchange realizes the connected vehicle [5], which is able to communicate with other vehicles, roadside units (RSU), pedestrians, and networks. This communication paradigm is called Vehicle-to-Everything (V2X) communication [1, 2]. To realize V2X communication, two incompatible communication standards are currently being used, namely (i) Wi-Fi-based and (ii) cellular-based V2X. Each of them uses specific communication technologies with their own capabilities, protocol stacks, and supported use cases [3–5]. The Wi-Fi-based approaches are divided into geographically specific implementations, which in turn are incompatible with each other.

Overall, new vehicular use cases such as remote driving, valet parking, platooning, and others envision the smart city. In this paradigm, connected autonomous vehicles and smart infrastructure interact with each other, streamlining traffic flow and reducing the impact on citizens [6]. This vision requires networking of many nodes and cyclic communication between them. Due to the functional complexity, this leads to increased V2X communication requirements in terms of bandwidth, transmission speed and transmission reliability. Therefore, it is important to know the performance of approaches in different scenarios. In this way, the most efficient approach can be selected for each specific use case to ensure the most resource-efficient data transmission.

Within the scope of this contribution our main objectives are:
- (O.1) the detailed analysis of the state-of-the-art V2X technologies and their performance
- (O.2) the analysis of the market penetration of existing V2X approaches Dedicated Short-Range Communication (DSRC Intelligent Transportation System G5 (ITS-G5), Cellular- and New Radio- Vehicle-to-Everything (C-V2X and NR-V2X)
- (O.3) the summary of existing challenges for a deployment of V2X in large scale real-world scenarios

The rest of the paper is organized as follow. In section II, the related work is summarized, whereas section III introduces the methodology of this paper. The sections IV and V are dealing with the summary of the technical basics regarding the existing V2X approaches and their performance. Furthermore, as there are several approaches used worldwide, their geographical usage is analyzed in section VI and the trend for their usage

*Corresponding author: Falk Dettinger*

Falk Dettinger, University of Stuttgart, Institute of Industrial Automation and Software Engineering (IAS), 70569 Stuttgart, Germany (e-mail: falk.dettinger@ias.uni-stuttgart.de).

Matthias Weiß, University of Stuttgart, Institute of Industrial Automation and Software Engineering (IAS), 70569 Stuttgart, Germany (e-mail: matthias.weiss@ias.uni-stuttgart.de).

Daniel Dittler, University of Stuttgart, Institute of Industrial Automation and Software Engineering (IAS), 70569 Stuttgart, Germany (e-mail: daniel.dittler@ias.uni-stuttgart.de).

Johannes Stümpfle, University of Stuttgart, Institute of Industrial Automation and Software Engineering (IAS), 70569 Stuttgart, Germany (e-mail: johannes.stuempfle@ias.uni-stuttgart.de).

Maurice Artelt, University of Stuttgart, Institute of Industrial Automation and Software Engineering (IAS), 70569 Stuttgart, Germany (e-mail: maurice-paul.artelt@ias.uni-stuttgart.de).

Michael Weyrich, University of Stuttgart, Institute of Industrial Automation and Software Engineering (IAS), 70569 Stuttgart, Germany (e-mail: michael.weyrich@ias.uni-stuttgart.de).

Mentions of supplemental materials and animal/human rights statements can be included here.

Color versions of one or more of the figures in this article are available online at http://ieeexplore.ieee.org

until 2030 is derived. In section VII, the currently existing challenges regarding V2X applications in large scale real-world scenarios are outlined. Finally, our results are discussed in section VIII and the conclusion is drawn in section VIII.

## II. RELATED WORK

In recent years, a large number of publications on Vehicle-to-Everything communication have been published, and surveys have been conducted to analyze and compare the content of the publications. In [7–11], the authors examine methods of cybersecurity and privacy issues in the context of V2X communication. In addition to common approaches [17], vulnerabilities in standards [16] and possible attack vectors [15] are analyzed, and challenges and solutions are presented.

The existing standards and challenges related to V2X communication are analyzed in [26, 27]. While Khezri et al. [26] deal with the benefits and barriers of V2X communication, Gyawali et al. focus on cellular-based V2X communication C-V2X and NR-V2X [27]. On the one hand, they focus on the status of the V2X cellular standard, and on the other hand, they describe specific challenges and their possible solutions in detail.

The surveys [25, 28, 29] examine the standards and use cases. In [29], Soto et al. describe cellular V2X standards and derive communication types and applications in the context of connected vehicles. In contrast, Alalewi et al. in [25] limit themselves to NR-V2X use cases, where the authors describe the use case requirements. In [28], Wang et al. also investigate V2X applications, but also focuses on their possibilities for testing and evaluation. In contrast, [30–32] discuss existing approaches to resource allocation in V2X communication.

Looking at the overview of the analyzed surveys, it is evident that there is already a strong focus on investigating and comparing security mechanisms and threats with respect to cybersecurity. V2X standards and resource allocation as well as applications and use cases have also been the subject of various analyses. However, to the authors' knowledge, a detailed comparison of the performance of existing V2X approaches and an analysis of the expected geographic deployment of the approaches has not yet been performed. It is therefore the aim of this paper to address this research gap.

## III. METHODOLOGY

In the following, the basic information about the considered review approaches will be summarized and the search strings as well as the selection procedure of the relevant articles and papers will be presented. Overall, the methodology used in this article was carried out in two stages:

First, to define appropriate search terms for the literature review, an initial keyword search was conducted. For this purpose, the relevant terms and topics of 15 publications with more than 30 citations in Google Scholar, dealing with *vehicle-to-everything communication* and *trends*, *cooperative* and *collective perception*, and *connected vehicles* were analyzed. The results of the initial keyword search, presented in Table 1, show heterogeneous cross-domain keywords, resulting in unspecific search strings and thus a large number of publications to be considered during the review.

During the literature review, a publication is considered if relevant terms are mentioned in the title, abstract, or keywords. To focus on the performance analysis and technical trends on current events, only literature from 2017 onwards is considered, since there is a higher number of active field tests involving V2X approaches starting in 2018. The *Web of Science*, *Scopus*, and *ACM* databases are used for the publication search because they are more focused than Google Scholar. *IEEExplore* is not explicitly considered, as its publications are directly synchronized with and included in the *Scopus* and *Web of Science* databases [33].

Based on Table 1, we consider publications dealing with V2X and V2X-related approaches based on IEEE 802.11 and cellular V2X and their performance analysis and global geographic trends during the literature review. This includes *DSRC*, *ITS-G5*, and *C-V2X*, *LTE V2X*, and *NR V2X* and their performance analysis. To outline geographical trends, the use of V2X in the *United States*, *Europe*, and *China* will be highlighted. From the overall literature, the current challenges in V2X communications will be derived.

The summarized search terms are shown in Table 2 along with the number of publications listed in the considered databases. Since all V2X and V2X-related approaches are relevant for this publication, the keywords V2X, *IEEE 802.11*, *DSRC*, *ITS-G5*, *C-V2X*, *LTE-V2X*, and *NR-V2X* are considered alternatively. To focus the results of the literature review, three subtasks are defined to search for technical fundamentals, performance evaluation, and geographical trends. As this contribution focuses on the vehicle domain, publications focusing on the rail, maritime, and air domains are excluded. In addition, to focus more on V2X communication, publications with specific frameworks and topics related to cooperative perception, such as map generation or point clouds, as well as safety and security aspects are excluded.

Table 1: results of the keyword search considering relevant terms and topics in [3, 12–25]

| Relevant terms | |
|---|---|
| Top-Level | Sub-Level |
| **Vehicle-to-Everything (V2X)** | |
| | IEEE 802.11p, bd |
| | ITS-G5 |
| | DSRC |
| | C-V2X |
| | LTE-V2X |
| | NR-V2X |
| | V2X-Messages |
| | Performance analysis of V2X approaches |
| | Approach related overview about technical basics |
| **Cooperative perception / Collective perception** | |
| **Trend** | |
| | Europe |
| | China |
| | United States |
| **Standards** | |
| | ETSI |
| | 3GPP |
| | Cooperative intelligent transportation systems (C-ITS) |
| | Connected and autonomous vehicles (CAV) |

Table 2: Search strings and number of filtered publications for the initial keyword search with Google Scholar and the literature review with the databases ACM, Web of Science (WoS) and Scopus.

| Topic | Search String | ACM | WoS | Scopus | Scholar |
|---|---|---|---|---|---|
| Initial keyword search (2019 - 2023) | "Vehicle-To-Everything" OR V2X" OR "("V2X" AND "trends" OR ((("cooperative" OR "collective") AND perception") OR "connected vehicles" | - | - | - | 8,330 |
| Technical basics to V2X-related approaches (2017 - 2023) | ("V2X" OR "Vehicle to Everything" OR "NR V2X" OR "C V2X" OR "DSRC" OR "ITS G5" OR "IEEE 802.11p" OR "LTE V2X") AND ("overview" OR "technical" OR "basics") NOT ("safety" OR "security" or "map" OR "Framework" OR "aerial" OR "rail") | 6 | 110 | 226 | - |
| Performance evaluation of V2X (2017 - 2023) | ("V2X" OR "Vehicle to Everything" OR "NR V2X" OR "C V2X" OR "DSRC" OR "ITS G5" OR "IEEE 802.11p" OR "LTE V2X") AND ("Performance" OR "Evaluation") NOT ("safety" OR "security" or "map" OR "Framework" OR "aerial" OR "rail") | 30 | 1,897 | 2,144 | - |
| Geographical technological trends in V2X (2017 - 2023) | ("V2X" OR "Vehicle to Everything" OR "NR V2X" OR "C V2X" OR "DSRC" OR "ITS G5" OR "IEEE 802.11p" OR "LTE V2X") AND ("trends" OR "Europe" OR "United States" OR "US" OR "China") NOT ("safety" OR "security" or "map" OR "Framework" OR "aerial" OR "rail") | 16 | 855 | 161 | - |

Searching with, e.g., *Web of Science*, shows 2,054 publications dealing with V2X performance evaluation excluding rail, maritime, safety and security applications and frameworks. 172 are summarizing an overview about V2X and its sub-approaches, while 1,125 address trends regarding V2X or geographical trends in China, Europe, or the United States. Because of the large number of publications, we focus on a semi-structured literature review [34, 35].

An initial semi-structured literature review is conducted to summarize the state-of-the-art in section III and IV. To analyze the performance of different V2X approaches, derive the worldwide usage of these approaches beyond 2030 and to show existing challenges in section V to VII, a meta-analysis is practiced. Both, the semi-structure review and the meta-analysis will consider the same literature basis, as most of the publications introduce technical basics and a performance evaluation regarding the approaches.

## IV. TECHNICAL BASICS

Technological progress in automotive applications is steadily evolving. Advanced driver assisted systems (ADAS) as well as semi-autonomous systems increase both vehicles drivability and on-road safety. Due to this intelligent and autonomous behavior, these systems are called intelligent transportation systems (ITS). Currently the usage of Level 3 autonomous systems like the Mercedes Benz drive pilot is allowed in scenarios with lower or limited complexity, such as on highways [36].

The interconnection of road participants enables the extension of ITS to cooperative ITS (C-ITS). They are expected to be an indispensable step towards the development and the establishment of autonomous vehicles [36]. Thus, besides vehicles, pedestrians and roadside units are also considered as part of C-ITS. Due to a high dynamic environment, high travel speeds and unexpected behavior of traffic participants, the interconnection between traffic participants requires three specific needs. Besides a low latency for data transmission, a high packet delivery ratio is needed, while maintaining the available maximum bandwidth of the communication channel. The related communication mechanism is called Vehicle-to-Everything communication.

Currently, there are two different approaches for V2X communication to enable the transmission of data and information between road users. Dedicated Short Range Communication (DSRC) is standardized by the United States of America, while Intelligent Transportation System G5 (ITS-G5) is standardized by the European Union. Both are based on the Wireless Local Area Network (WLAN) standard, IEEE 802.11p (i), but use a different communication stack. Cellular V2X (C-V2X) and New Radio V2X (NR-V2X) (ii) use the 4th (Long Term Evolution (LTE)) and 5th generation (5G) cellular standards [4, 37]. In the following sections, the technical basics and the development timeline of both V2X communication approaches are presented. For this purpose, the development and associated communication stacks of DSRC and ITS-G5 as part of the IEEE 802.11 standard are presented in Section A. C-V2X and NR-V2X are then introduced in Section B.

### A. Wi-Fi based V2X

Both DSRC and ITS-G5 are based on the IEEE 802.11p standard [38]. This is a modified WLAN protocol introduced in 2010, in which changes were made to the physical layer (PHY) and the data link layer, more precisely the MAC (Medium Access Control) layer [3]. The initial requirements for IEEE 802.11p were a relative vehicle velocity of 200 km/h, a response time of 100 ms and a communication range up to 100 m [39].

At the PHY layer, the operating frequency of the 802.11p standard has been increased from 5 GHz to 5.9 GHz compared to the 802.11a standard [40]. Other changes were made to the parameterization of the Orthogonal Frequency Division Multiplexing (OFDM) modulation method [41]. The IEEE 802.11p standard uses Binary Convolutional Coding for error correction. However, its ability to recover erroneous messages is lost as soon as higher Modulation Coding Schemes (MCS) are used for higher bit rates at distances greater than 50m [41].

More fundamental changes have been made to the MAC layer. In general, the IEEE 802.11p standard defines a Basic Service Set (BSS) for network construction. This is a group of network devices that allows the use of different network topologies, such as mesh networks or access points. In order for a device to send messages within a device group, it must be a member of that BSS. However, membership involves procedures such as channel scanning, association, and authentication. The time required for this is not available in the V2X context [42, 43], so an alternative mode called Outside the Context of BSS (OCB) has been defined in IEEE 802.11p. In this mode, the BSS control procedures are omitted, resulting in a lower time overhead for message transmission. OCB is particularly suitable for fast transmission of short messages. Network devices operating in OCB mode use Enhanced Distributed Channel Access to access the transmission medium. This is based on the Distributed Coordination Function (DCF), which manages the access of multiple nodes to a wireless network. In IEEE 802.11p, the access method used is CSMA/CA (Carrier Sense Multiple Access with Collision Avoidance) [44].

In CSMA/CA, the participants listen to the communication medium [40]. If the medium is free for a certain time, the station willing to transmit starts the transmission. If the channel is busy, the transmission is delayed by a random period (Contention Window, CW). Prioritization of messages is made possible by assigning Access Categories (ACs) depending on the content of the message

Because IEEE 802.11p does not meet the high latency and reliability requirements of new Day2 and Day3+ use cases, the IEEE extended the 802.11p standard to 802.11bd in 2018 [47]. In doing so, the requirements for the standard were expanded and technical improvements introduced and tested in the 802.11n/ac/ax standards were adopted in 802.11p with the primary goal of achieving higher throughput and greater resilience. The requirements of the IEEE 802.11bd standard have also been adapted. For example, it requires a relative speed of 500 km/h, a mode with at least twice the range of IEEE 802.11p, and at least one mode for localization. In addition, 802.11bd must also support 802.11p devices and be compatible with them at the communication layer [39, 47]. Among other things, to increase performance, the following modifications were made [47, 48] to increase to a factor of 3, and to improve robustness and interoperability:

- Binary Phase-Shift Keying- Dual Subcarrier Modulation (BPSK-DCM) and 256 QAM as well as Multiple Input Multiple Output operation in unicast mode have been added to the modulation schemes.
- In addition, 52 data subcarriers are now supported instead of 48 data subcarriers in 802.11p.
- In addition to BCC, Low-Density Parity-Check (LDPC) is also supported for error correction.
- The frequency band has been extended to the 60 GHz band and localization is supported to enable new high-performance applications.
- One to three retransmissions of messages are also supported.

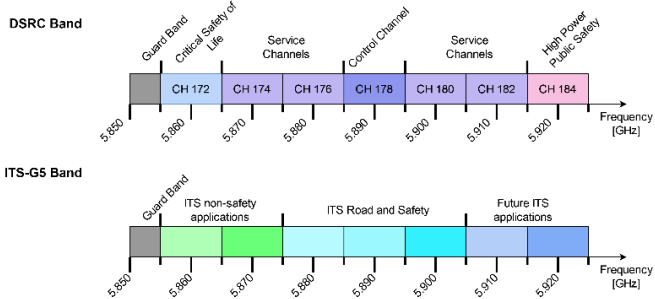

**Figure 1: DSRC (top) and ITS-G5 (bottom) band Channel frequencies according to [40, 45, 46]. For the ITS-G5 Band, no channel names are available.**

Currently there exist two specific implementations, namely DSRC and ITS-G5. DSRC was developed in the USA, while ITS-G5 was standardized by the European Telecommunications Standards Institute (ETSI) under ESI EN 302663 [38, 42]. For the sake of completeness, it should be mentioned that ITS-Connect was standardized as the Japanese version of DSRC/ITS-G5 by the Association of Radio Industries and Businesses (ARIB) [2]. As mentioned at the beginning, DSRC and ITS-G5 are based on the IEEE 802.11p standard, but the approaches differ in some respects [2, 40, 49]. In the USA [3], DSRC has been mandatory for new vehicles since 2016, while in the EU [50] both DSRC and ITS-G5 are de facto standards. Here, the first V2X-capable vehicles were produced in 2019.

*Dedicated Short Range Communication (DSRC)*

In 1999, the Federal Communications Commission (FCC) reserved a 75 MHz frequency band in the 5.820 GHz and 5.925 GHz range for automotive use [51]. In 2002 [49], the American Society for Testing and Materials (ASTM) published ASTM E2213, which proposed the use of a modified IEEE 802.11a standard to implement DSRC. This was taken up by the IEEE and the IEEE 802.11p standard was created. The transmission range is in the range of a few hundred meters compared to classic mobile radio technology [40].

The frequency band provided by the FCC is divided into seven channels [2], each 10 MHz wide. One channel is designed as a control channel (CCH), while the other six channels are used for service delivery (SCH). The remaining 5 MHz, which is not assigned to any channel, is used as a guard band to prevent interference (Figure 1 (top)).

While the CCH is intended for broadcast transmission of safety-related messages, messages for infotainment and traffic purposes are transmitted via the SCH [2]. In addition to the IEEE 802.11p standard, the DSRC communication stack also includes the IEEE 1609.x standard, which together form the main part of the Wireless Access for Vehicular Environment (WAVE) stack [40]. However, the term IEEE 802.11p is usually used to refer to the entire WAVE protocol stack [51], as shown in Figure 2 (IEEE WAVE).

The IEEE 1609 standard [51] is used for direct communication between vehicles and between vehicles and the traffic infrastructure. In this context, IEEE 1609.2 implements digital security for message transmission. It provides authentication mechanisms and message encryption using digital signatures and certificates. Certificates do not contain driver-specific information and are changed cyclically, protecting the identity of the driver. IEEE 1609.3 defines WSMP (Wave Short Message Protocol) [40], which is a single-hop network protocol with a small header and supports message multiplexing, where messages are only transmitted to communication partners one hop away from the current access point. The MAC sublayer extension allows DSRC systems to efficiently switch between communication channels when there are multiple transceivers.

*Intelligent Transportation System G5 (ITS-G5)*

The ITS-G5 standard also operates in a 75 MHz spectrum that is divided into seven 10 MHz channels [46]. However, in contrast to DSRC, the frequency band is additionally divided into functional groups. The band from 5.875 GHz to 5.905 GHz corresponds to the main frequency band and is used for the transmission of safety-related messages and traffic efficiency messages. For non-safety related messages, the 5.855 GHz to 5.875 GHz frequency band is used. The frequency range from 5.905 GHz to 5.925 GHz is reserved for future expansion [46]. Figure 1 (bottom) shows the frequency bands structure.
Similar to DSRC, ITS-G5 uses the IEEE 802.11p standard in the PHY and MAC layers [53]. The structure of the ITS-G5 stack therefore only differs in the layers above, which is why ITS-G5 supports additional features compared to the WAVE components. The structure of the protocol stack is shown in Figure 2 (ETSI ITS-G5). In order to limit channel occupancy and to give priority messages a high delivery probability, the ITS-G5 standard also uses Decentralized Congestion Control (DCC) methods [2]. The GeoNetwork (GeoNet) protocol is used in the transport layer. This is a standard defined by ETSI EN 302 636 that allows ad hoc multi-hop communication between nodes. It allows addressing and forwarding of messages using geographical coordinates. Therefore, messages can be sent and forwarded even if a communication partner is not directly in the communication range of the sender. Compared to the WAVE Short Message Protocol (WSMP), multi-hop communication via geo-addressing is more complex and has a higher overhead. On the other hand, the Basic Transport Protocol (BTP) is used to realize the end-to-end transport of messages to the ITS devices.

*B. Long Term Evolution (LTE)/5G based V2X*

With the advent of more powerful wireless communication standards, C-V2X (Cellular V2X) has been developed based on 4G and 5G communication [4, 37]. This is divided into LTE-V2X, which uses LTE (Long Term Evolution) communication, and NR-V2X, which uses 5G-NR (New Radio) communication. C-V2X is intended to enable new use cases and applications compared to IEEE technology. Unlike DSRC, for example, which is only licensed in the 5.9 GHz band, C-V2X communication is also supported in the licensed sub-6 GHz band. In contrast to the 802.11 standard, the standardization is done by the 3GPP (3rd Generation Partnership Project), which is responsible for the general standardization of mobile radio. [54]. LTE was defined in 2008 with Release 8 (Rel-8) and has been enhanced with new features in recent years.

In C-V2X Release 12, also known as LTE-Advanced, the D2D (Device to Device), also known as Sidelink (SL) communication, was introduced [1, 37, 55] using the PC5 interface to support Proximity Services (ProSe) without the need of the base station (eNodeB, eNB) for short-range

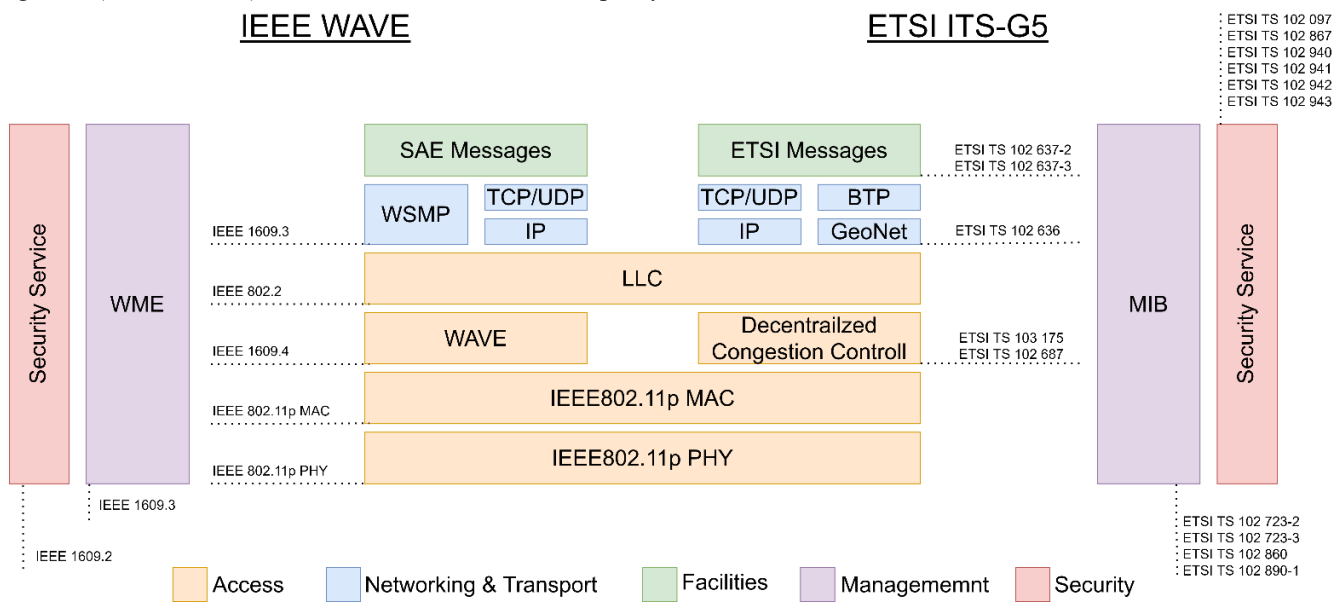

Figure 2: left: WAVE protocol stack considering the included sub-layers for security, MAC And PHY Layer according to [40, 52] and ITS-G5 Stack containing physical and MAC layer as well as overlying layers according to [49, 52] (right)

communication between two or more network nodes using the PC5 interface, through which V2V, V2I and V2P communication can be efficiently realized. On the other hand, the Uu interface enables communication between the user equipment (UE) and the base station (eNodeB or eNB), which is directly used for V2N communication. [56]. To enable the D2D communication, the eNB is bypassed by network nodes and thus, the data traffic via the eNB is reduced. In addition, D2D communication can be used to forward data when subscribers have no or poor connection to the eNB. Overall, two different modes [57] are defined for scheduling the message transmission of a network node (UE). Mode 1 is a centralized connection approach. Here, the planning of the dedicated radio resources of the individual UEs is done by the eNBs in coverage, therefor called *in-coverage mode*. In contrast, in Mode 2 the radio resources are selected independently by the UEs from a defined resource pool and the transmission is performed via the PC5 interface. In Mode 2, no UE needs to be within range of the subscribers, which is why this mode is also referred to as *out-of-coverage mode*. Due to the high communication latency of D2D communication in Rel-12, it was not suitable for V2X communication [57]. Figure 3 shows the 3GPP release history for C-V2X technology.

Building on the D2D communication in Rel-12, LTE-V2X communication was specified in Rel-14 [2]. This is based on the time-division LTE (TD-LTE) based LTE-V presented by the China Academy of Telecommunication Technology (CATT) in 2013. Similar to C-V2X, this includes two modes for centralized and decentralized data transmission and was standardized by 3GPP as LTE-V2X in late 2016 [37]. LTE-V2X is specifically designed for use in the automotive environments [58]. Since the publication of LTE-V2X, the development of C-V2X communication has been divided into phases. Rel-14 marks the beginning of phase 1, which was completed in March 2017 [37].

With Release 15 in 2017, Phase 2 of C-V2X was initiated, introducing several improvements and enhancements to the LTE V2X technology were introduced. These include support for advanced V2X scenarios such as platooning or remote vehicle control [3]. Performance improvements have also been introduced. For example, 64-QAM was introduced as an additional coding scheme [37], which increases the number of transmitted bits per symbol and thus the data rate compared to the originally used 16-QAM improving the performance of the PC5 interface.

Like the LTE uplink [1], LTE-V2X uses the single-carrier frequency-division multiple access (SC FDMA) method in the PHY layer. The channel width is either 10 MHz or 20 MHz and is divided into resource blocks (RBs). The overall structure of LTE radio frames is summarized in Figure 4. These have a width of 180 kHz and represent the smallest unit that can be allocated to a UE as a resource in the frequency domain. Each of these RBs contains 12 subcarriers, each with a bandwidth of 15 kHz. In the time domain, the channel is divided into subframes of 1 ms length, which are called Transmission Time Interval (TTI), which is the smallest unit in the time domain.

During a TTI, 14 OFDM symbols are transmitted, nine of which are used for data transmission and four for the transmission of Demodulation Reference Signals (DMRSs) [5]. The DMRSs are used to estimate the channel parameters and to suppress Doppler effects caused by the high vehicle speed. The remaining OFDM symbol is used as a guard symbol [1].

A group of RBs in the same subframe is called a subchannel and the number of RBs per subchannel can vary depending on the configuration and the amount of data [59]. The subchannels are used to transmit data and control information, where the data is organized into Transport Blocks (TBs) that are sent over the Physical Sidelink Shared Channel (PSSCH). A TB contains a complete message packet, such as a CAM or BSM, and occupies a different number of subchannels depending on the number of packets sent. In addition to the data, a corresponding Sidelink Control Information (SCI) message is sent per TB over the Physical Sidelink Control Channel (PSCCH) assigned to the TB, which requires two RBs. This message contains information such as the type of modulation, the priority of the message, and whether it is an initial transmission or a blind retransmission of a message. If a TB requires retransmission, the same message structure is sent in a different subframe [57]. In this case, the receiver uses the Hybrid Automatic Repeat Request (HARQ) procedure to combine the two identical TBs from transmissions one and two to form a resulting message. The repeated sending of a message by the sending UE is done without prior knowledge of whether the message was successfully received the first time. This means that the sending UE performs the resending blindly [4]. Overall, this leads to an increased reliability of message delivery.

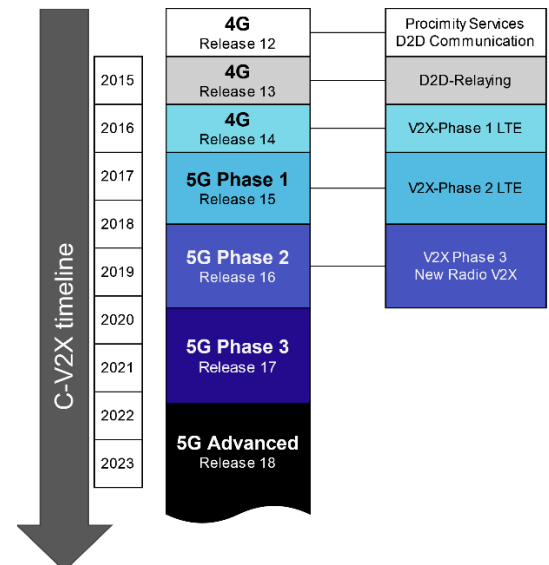

**Figure 3:** V2X-Timeline starting in 2014 with Rel-12, till the end of 2023 with Rel-18 according to [54, 60, 61]. Timeline shown on the left, state of 4G and 5G radio communication (center) and V2X state (right).

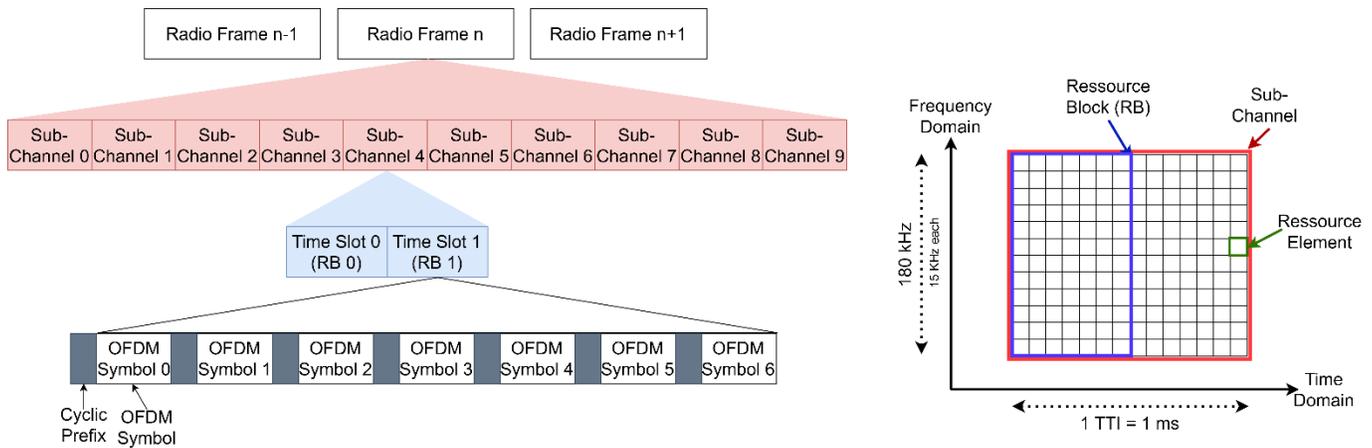

Figure 4: left: Structure of Radio frames in LTE-V2X. In LTE each radio frame contains out of ten sub-frames, each 1 ms long while each sub-frame consists out of two timeslots with maximum seven OFDM symbols per slot. Right: Structure of one time slot also named sub-channel. These contain of two resource blocks (RB) with 12 carrier signals each 15 kHz wide and 0.5 seconds long. [1, 5, 41]

LTE-V2X introduces two new modes for C-V2X communication, Mode 3 and Mode 4 [37, 62]. These are in addition to the Mode 1 and Mode 2 modes introduced in Rel. 12. In Mode 3, similar to Mode 1 of the initial C-V2X, resource management is controlled by an eNB. The UE requests resources from the eNB via the LTE-U (Uu) interface. The PC5 interface is then used for communication between the nodes. In Mode 4, similar to NR V2X Mode 2, resource management is not performed by the eNB but directly by the individual UEs. Communication then takes place directly via the PC5 interface. Al though the operation of modes 1/3 and 2/4 is similar, they differ significantly in their structure and the way PSCCH and PSSCH are allocated [44]. Mode 3 specifies two scheduling approaches for resource selection. In dynamic scheduling, UEs request subchannels from the eNB for each TB. In contrast, in Semi Persistent Scheduling (SPS), the eNB reserves the subchannels for the UEs. In Mode 4, the Sensing Based Semi Persistent Scheduling (SP-SPS) introduced in Rel-14/15 is used. The UEs then select their own subchannels. In Mode 3, higher performance can be achieved through centralized resource allocation, although there is a higher overhead due to communication with the eNB [5].

For a low density of communication participants, the defined requirements for V2X communication via LTE-V2X can be achieved in terms of latency and bandwidth for simple data transmission [57]. However, the Quality of Service (QoS) requirements are constantly increasing. The performance limits of LTE-V2X have been exceeded, in particular, by enabling new use cases such as platooning or extended sensors, as described in section. III.D, the performance limits of LTE-V2X have been exceeded. the solution is a more powerful standard called New Radio (NR)-V2X. This is based on 5G New Radio (NR) technology. 5G NR was introduced by the 3GPP in V2X Rel-15, initially without sidelink communication [5, 57]. Based on 5G NR, NR-V2X was then introduced in Rel-16, including the sidelink for D2D communication. The development of Rel-16 started in June 2018 and was completed in July 2020 [4, 57]. With this progress, the V2X Phase 3 has been completed.

In 2020, the content extension for 5G Release 17 was started and the release was frozen in 2022 [61]. The focus was on enhancing existing features such as sidelink communication, power saving, and coverage and positioning improvements [60, 63]. In addition, some new features are provided, with a focus on the integration of satellite access within the 5G network. The changes are summarized in TR 21.917, which is not finally published in December 2023. Since Q4 2021, 3GPP has been working on 5G "Advanced" in Release 18, which is expected to be released in 2024 [63]. The focus is on system architecture and related services as well as radio layers and interfaces [64].

Therefore NR-V2X is not supposed to replace LTE-V2X [5], rather than to extend it. As the 5G technology is more powerful than 4G (LTE), classic use cases like the transmission of safety related messages like CAMs as well as enhanced V2X (eV2X) – sometimes also called advanced V2X – use cases can be covered. 25 new use cases, assigned to four groups were defined in NR-V2X [37]. To support these new use cases two additional cast types, called unicast and groupcast were introduced, in NR-V2X [57]. LTE-V2X only supports broadcasting, where all nodes within the communication range of the sender can receive the message. In contrast, unicast allows communication between two dedicated UEs, while groupcast allows the message transmission from a UE to a specific group of subscribers. In NR-V2X, a UE can communicate via several cast types simultaneously using the PC5 interface [54].

The NR-V2X SL in Rel-16 is designed for frequency range 1 [5]. This starts at 0.41 GHz and ends at 7.125 GHz. In addition, the 5.9 GHz band, which is also used by LTE-V2X and DSRC, and the 2.5 GHz band for WLAN can be used. Frequency range 2 (24.25 GHz to 52.6 GHz band) is also supported, but was not the focus of Rel-16.

As introduced, the NR-V2X supports a variety of different communication types such as unicast, groupcast and broadcast. This provides greater flexibility and enables new use cases to be realized. In NR-V2X, the modulation scheme of C-V2X in Rel-15 is extended by the 16-QAM of Rel-14. This results in backward compatibility. In addition, the NR-SL supports 256-QAM [65], which enables a higher throughput over the sidelink channel. Since most V2X use cases require a communication distance of less than 1000 m, the connection speed gains higher priority. Thus, the change from SC-FDMA to OFDM was implemented, which increases the connection speed while reducing the coverage at the same time. To increase the reliability for the group and unicast communication [39], a feedback channel is provided on the sidelink channel. This eliminates the need for blind retransmissions by the sender, as the receiver confirms the correctness of the transmission. To reduce the latency the time division multiplexing is used in NR V2X. The receiver obtains all necessary information for decoding, followed by the actual data. Additionally, NR-V2X supports a flexible frame structure with subcarrier spaces between 15 kHz and 480 kHz [1]. This leads to shorter transmission intervals (TTI) and explicitly reduces latency.

Furthermore two new modes for resource allocation were introduced with NR V2X. Namely SL Mode 1 and SL Mode 2 [37]. SL Mode 1 is equal to the LTE-V2X Mode 3, where the so called gNB (instead of eNB in Mode 3) is responsible for the resource scheduling. In SL Mode 2 (similar to LTE-V2X Mode 4) UEs can communicate with each other directly using the sidelink channel.

### C. Coexistence between LTE-V2X and NR-V2X

As introduced in the beginning, NR-V2X should not replace LTE-V2X, rather than extend it. So, with LTE-V2X and NR-V2X different concepts are available [57]. NR-V2X is not compatible with LTE-V2X due to the usage of different numerologies like scalable Transmission Time Interval (TTI) durations and sub-carrier spacings (SCS) in cases where the higher performance of NR-V2X is required. Therefore, new vehicles will likely be equipped with both radio access technologies [39].

While LTE-V2X covers classic use cases, NR-V2X is used for advanced ones [5]. It is intended that vehicles select the appropriate Radio Access Technology (RAT) based on the current use case. To enable the coexistence of LTE-V2X and NR-V2X, mechanisms have been defined in Rel-16, also known as In-Device Coexistence. These are responsible for managing SL resources at the vehicle and device level. Furthermore, the coexistence of both RATs at the network level is defined [39]. This mechanism is referred to as cross-RAT control. In Rel-16, it is only applicable when LTE-V2X and NR-V2X operate on different channels. The challenges of in-device coexistence lie in the management of resources such as used channels and radio frequencies while both RATs are in use [5]. The main challenges are the limited transmission power as well as possible interferences between the RATs due to the simultaneous use of non-well separated frequencies. Rel-16 defined two possible solutions to mitigate interference between LTE and NR-V2X, namely the use of Time Division Multiplexing (TDM) and Frequency Division Multiplexing (FDM).

With the TDM approach [5], only one RAT can access the transmitter at a time. Therefore, only alternate transmission of messages is possible and further synchronization is required.

With FDM [5], both RATs can be used simultaneously and the available transmission capability is shared between the RATs and can be allocated statically or dynamically based on the message priority.

### D. Use-Cases

The 3GPP defines C-V2X and LTE-V2X use cases (in TR 22 885) based on the ETSI (ETSI TR 102 638). These groups of use cases require a maximum latency of 100 ms and packet delivery ratio of 95 % [19]. Thereby the requirements for different use cases are not identical. Some use cases require lower latency, other higher reliability, or bandwidth.

With the more advanced communication technique of NR-V2X, 25 new use cases were defined by the 3GPP. They are divided into four groups namely Vehicle Platooning, advanced Driving, Extended Sensors and Remote Driving [5, 66–68].

As shown in Table 3 there are three relevant parameters that need to be considered in different use cases. Latency describes the time between sending a message to a sender, processing it at the receiver, and receiving the results at the original sender. The use case latency depends on the distance between the vehicles and their speed, while the system latency depends on the network structure and the current network situation such as traffic and number of participants. In particular, lower latency can only be achieved with SL communication, as multi-hop communication increases latency due to routing. The packet delivery ratio describes how many messages sent from a source to a destination must be delivered correctly. It depends on the distance between vehicles, the maximum communication distance of the RATs, and the environmental conditions. Bandwidth requirements are described by the data rate and define the number of bits transmitted per second. The required data rate is directly related to the number of messages and their payload. Often, the data rate differs between the V2V data rate using the SL channel and the V2I data rate and the V2N data rate.

Vehicle Platooning: Vehicles in motion form a dynamic convoy. Within a convoy, vehicles communicate by sending periodic messages to ensure the convoy's functionality. The goal is to optimize transportation and fuel efficiency. This requires latency between 10 ms and 500 ms, depending on the speed and number of vehicles in the convoy, and a message delivery rate in the range of 90 to 99.99 %. As vehicles dynamically join or leave the convoy, maneuvering within the convoy and with road infrastructure or pedestrians needs to be coordinated.

Advanced Driving: Vehicles share sensor information and maneuver intentions to negotiate their trajectories. As vehicles travel at different speeds, low latency data transmission is required, ranging from 3 ms to 100 ms depending on the situation and speed. In addition, to ensure safe maneuver coordination, a high packet delivery ration is required, ranging from 90% to 99.99%. Therefore, 0.5 Mbps to 50 Mbps is required for V2V and V2I communication, as status information and planned maneuvers need to be distributed among different participants.

Table 3: ETSI Use Cases and their requirements regarding latency, reliability and bandwidth [66–68]

| V2X Use case | Latency (ms) | Reliability (%) | V2N Data rate (Mbps) | V2V Data rate (Mbps) | V2I Data rate (Mbps) |
|---|---|---|---|---|---|
| Platooning | 10 - 500 | 90 - 99.99 | 80 - 350 | 2 - 65 | 2 - 50 |
| Advanced Driving | 3 - 100 | 90 - 99.99 | 10 - 50 | 0.5 - 50 | 0.5 - 50 |
| Extended Sensors | 3 - 50 | 95 - 99.99 | 10 - 50 (Map Sharing) | 25 - 1000 (collective Perception) | - |
| Remote Driving | 5 | 99.999 | DL: 1 UL: 25 | - | DL: 1 UL: 25 |

Extended Sensors: Enables the exchange of sensor data with nearby vehicles, RSUs, pedestrian devices, and V2X servers. This includes pre-processed data. The goal is to increase the awareness of individual road users and to extend the field of view of the participants. Due to the variable speed and highly dynamic environment, a low latency between 3 ms and 50 ms is required. Due to the dynamic nature of the environment, a packet delivery ratio between 95% and 99.99% is required. Depending on the level of data distribution, data rates between 10 and 50 Mbps are required for collaborative map sharing between vehicles and networks, and 25 Mbps - 1000 Mbps for collective perception between different vehicles.

Remote Driving: Enables the remote control of vehicles. This includes both driving without passengers and transporting passengers. Another important use case is the remote control of automated vehicles that are in a situation where they can no longer control themselves. This requires a low latency of about 5 ms and a high reliability of 99.999% to enable safe remote control of a vehicle. The stream of sensor data from the vehicle, as well as infrastructure status information, must be transmitted to the remote driver. This results in an average uplink of approximately 25 Mbps. Since only the steering signals are transmitted to the vehicle, the downlink is about 1 Mbps. Table 3 summarizes different parameter values based on different literature references [66–68].

## V. Performance Analysis

This section provides an overview of several different research efforts of recent years that have focused on the performance of LTE-V2 and NR-V2X and the implementations of IEEE 802.11p, namely DSRC/ITS-G5. First, the performance metrics commonly used for comparison are briefly explained in Section V.A. Subsequently, the results of the individual works are summarized in Section V.B. Finally, an assessment of the two technologies based on previous results is made in Section V.C.

### A. Performance metrics

Within the studies considered, various metrics are used for performance assessment. These metrics and evaluation scenarios are briefly explained below.

In general, two different environments are considered for evaluation. For Line of Sight (LOS) measurements, two or more communication partners have a line of sight to each other that is not interrupted by objects. On the contrary, None-Line of Sight (NLOS) refers to two or more communication partners that cannot establish a line-of-sight connection because there are objects in between. Additionally, the papers consider different traffic density scenarios, i.e., if there are many or only few partners are involved.

Across the observed papers, the following metrics were identified as common and are thus used for direct comparison:
- Packet Reception Ratio (PRR): The ratio between the number of neighbors that correctly decode a transmitted message and the number of total neighbors within a given distance of the transmitting vehicle [1].
- End-to-End (E2E) Latency: the time required to transmit a message from the sender's application layer to the receiver's application layer. Thus, it includes channel access [2].
- Update Delay (UD): The time difference between two successive successfully decoded messages from a sender to a receiver located at a certain distance from the sender [56].
- Packet Delivery Ratio (PDR): The ratio between the number of correctly received messages at a receiver and the number of retransmitted messages from a reference transmitter.

Additionally, the following metrics are used by a partition of the papers and mentioned if measured:
- Packet Loss Rate: The percentage of data packets that are transmitted but fail to reach their destination, indicating the reliability of the communication system.
- Average Bit Rate: The average amount of data transferred per unit of time, measured in bits per second (bps), reflecting the data transmission speed.
- Average Latency: The average time taken for a data packet to travel from the sender to the receiver, indicating the responsiveness of the communication system.
- Average Range: The typical maximum distance over which communication can occur effectively without significant loss of signal quality.
- Channel Busy Ratio (CBR): The proportion of time the communication channel is occupied with transmissions, showing the channel's utilization and congestion level.
- Propagation and Collision Error: Errors in data transmission caused by signal propagation issues and collisions when multiple transmissions interfere with each other.
- Received Signal Strength Indicator (RSSI): A measure of the power level received by the receiver, indicating the strength and quality of the received signal.

### B. Survey Results

Direct performance comparison between V2X communication via IEEE 802.11p and LTE-V2X has long been a focus of research. An overview of studies comparing the performance of communication technologies using IEEE 802.11p standard and LTE-V2X with Mode 3 and 4 is given by A. Keyvan in [44]. The studies contained therein are from 2017 and 2018 and the technologies are compared in two different scenarios with high and low traffic density respectively. However, Keyvan does not explain how one technology is superior to the other, but he infers that neither technology is superior to the other in all aspects. He also points out that most of these studies are subject to limitations, such as being simulations under ideal conditions, and that while LTE-V2X performs better than IEEE 802.11p in large areas, the potential improvements of IEEE 802.11p have been under-studied.

The authors in [37] investigate the performance of LTE-V2V in modes 3 and 4 using simulation and consider the PDR as a function of distance in a highway scenario. They vary the traffic density (in vehicles per meter), packet transmission rate (in packets per second, pps), and transmit power. They observe a decrease in PDR in both modes when the packet transmission rate is increased, the traffic density increases, or the transmit power is decreased. However, the performance in mode 3 operation is always better than the performance in mode 4. Regardless of the operating mode used, the work shows that a PDR of over 90% can only be guaranteed at very short distances (< 100m). Metrics apart from PDR like average bit rate, average latency, CBR, and propagation and collision error are mentioned as future work considerations but are not specifically compared in the provided simulations.

In [69], Guizar et al. present a system for cooperative collision avoidance that involves the exchange of so-called "occupancy maps" over LTE-V2X. For performance evaluation, these 700-byte messages are exchanged both using V2I communication with an RSU and V2V communication in a simulation. A large intersection in the city serves as the scenario, and the three transmission modes (V2I uplink, V2I downlink, V2V) are considered separately in terms of their PDR. For the V2I uplink, a PDR of about 70 % and, for the V2I downlink, a PDR of almost 96 % could be measured. With a distance between the vehicles of up to 150 m, a PDR of almost 74 % was measured for V2V communication. For V2I downlink communications, the latency to retrieve information from the RSU was less than 120 ms. For V2I uplink, latency to retrieve Cooperative Awareness Messages (CAM) was 106 ms to 160 ms. For V2V communications, latency to share CAM was 103 ms to 140ms for 150 m range. CBR was indirectly addressed through the TTI (Transmission Time Interval) occupancy rate, which showed that MCS-15 (with 22 RBs per TTI) had fewer collisions and was more reliable under high vehicle densities. The authors conclude from these results that their proposed system would be feasible using LTE-V2X communication, provided that all UEs are interconnected.

Ulbel investigates ITS-G5 in real road traffic in [70], where she focuses on the aspects of maximum range and connection quality. In her test drives, ranges of up to 3 km could be achieved for message transmission when the RSUs were mounted at the same height as traffic lights. Considering the link quality (signal strength min. -80 dBm), ranges of 90m (NLOS) – 300m (LOS) can be achieved. However, the connection quality was disregarded in the measurements. Packet loss rate was significantly influenced by the line of sight. Up to a distance of about 800 meters, no packet drop was observed when the sender and receiver were in direct line of sight. However, the smallest obstructions led to drastic packet loss.

The study [71] by R. Molina-Masegosa and J. Gozalvez compares LTE-V2X and IEEE 802.11p technologies in terms of packet delivery ratio (PDR) at varying distances and transmission rates. LTE-V2X outperforms IEEE 802.11p at distances beyond 200m, especially at higher transmission rates. However, IEEE 802.11p performs better at shorter distances and lower transmission rates. LTE-V2X generally experiences a lower CBR than IEEE 802.11p under the same traffic density, especially when handling aperiodic messages. This is because packet collisions, which are more frequent in LTE-V2X, generate less channel load due to collided packets contributing less to the channel load.

In [58], K. Jellid and T. Mazri examine PDR versus traffic density for DSRC and LTE-V2X technologies in a highway scenario. LTE-V2X exhibits superior PDR compared to DSRC at shorter delivery times, lower latency and higher traffic densities. Unfortunately, the authors do not provide any information about the mode in which LTE-V2X is operated and the distance between the vehicles. LTE V2X offers a much higher bit rate (up to 1 Gbits/s) compared to DSRC (3-27 Mbits/s), indicating that LTE V2X can support more data-intensive applications. Additionally, simulation results indicate that LTE V2X consistently achieves lower latency than DSRC. For example, even with an increase in the number of vehicles, LTE V2X maintains a latency below 100 ms, while DSRC's performance degrades significantly under congestion.

T. Shimizu et al. compare DSRC and LTE-V2X PC5 Mode 4 in [43], analyzing different traffic density scenarios on a 12-lane highway. The traffic density varies in the simulation from 108 vehicles/km up to 423 vehicles/km at a speed of 100 km/h. LTE-V2X offers better PDR at longer distances, while DSRC demonstrates lower latency. This difference is explained by the two different media access methods (CSMA/CA for DSRC, Sensing-Based Semi Persistent Scheduling (SB-SPS) for LTE-V2X).

V. Maglogiannis et al. [72] conduct a field test comparing LTE-V2X and ITS-G5 technologies on a highway. LTE-V2X outperforms ITS-G5 in V2I communication at distances beyond 200m, while ITS-G5 shows lower latency. For instance, 97% of packets transmitted by ITS-G5 have latency smaller than 5.5 ms, while C-V2X PC5's latency for the same percentile is up to 37.84 ms. In V2V communication, LTE-V2X offers longer ranges than ITS-G5.

Results of another field test comparing IEEE 802.11p and LTE-V2X are presented by M. Shi et al. in [51]. In the field trial presented there, four different scenarios are considered, each with two vehicles. In the first scenario, both vehicles are at distances of 100 m and 800 m from each other, while in the second scenario, both vehicles follow each other at the same speed at a distance of about 100 m. In scenario three, the two vehicles travel at the same speed but in opposite directions, and in scenario four, the two vehicles travel different routes and then meet at an intersection at the same time. In this case, an NLOS situation also occurs in places. All scenarios thus represent situations with low traffic volumes. LTE-V2X demonstrates better PDR over distance, while 802.11p exhibits lower latency. LOS situations favor V2V communication for safety, although collisions can still occur in NLOS situations. In general, the paper indicates that both standards face significant challenges in NLOS scenarios, leading to higher propagation and collision errors. LTE-V2X shows better performance in maintaining communication under these conditions compared to 802.11p.

A. K. Gulia [2] evaluates the performance of the IEEE 802.11p standard and LTE-V2X in Mode 3 using a 4-lane highway scenario with a length of 1 km and a maximum speed of 30 km/h. Furthermore, he varies the vehicle density in four levels: 10, 30, 50 and 70 vehicles/lane/km. He calls these low, moderate, high, and extremely high traffic density, respectively. 300-byte Basic Service Messages (BSM) or CAMs are sent with a frequency of 10 Hz and 20 Hz. For the IEEE standard, the default data rate of 6 Mbit/s per channel is used. Metrics used include PRR and End-to-End latency. The results show that IEEE 802.11p is better than LTE-V2X Mode 3 at a transmission frequency of 10 Hz in almost all traffic densities. However, LTE-V2X Mode 3 has a better average PDR in extremely high traffic situations, even if the End-to-End latency suffers.

M. Karoui et al. [73] compare ITS-G5 with LTE-V2X in mode 3, taking into account LTE traffic caused by applications that are not related to V2X. Generally, ITS-G5 provides lower E2E delay, stable CBR, and consistent RSSI under varied conditions, making it more suitable for critical ITS services that require lower latency. LTE-V2X Performs well under ideal conditions but is more affected by concurrent non-ITS traffic, increasing density, and handover processes, leading to higher E2E delays and variable RSSI. LTE-V2X offers broader coverage but needs careful consideration of network load and handover impact for reliable ITS service performance. In their paper, the authors also provide an overview of other works comparing C-V2X with ITS-G5 or DSRC.

In [74], the IEEE 802.11p standard and LTE-V2V are compared in both modes. For this purpose, the performance metrics PDR and update delay (UD) are used. In the paper, the authors conclude that LTE-V2V in Mode 3 has better PDR than IEEE 802.11p and that LTE-V2V in Mode 4 has the worst Update Delay. In another study by A. Bazzi et al. [75] the authors also conclude that LTE-V2X Mode 3 is superior to both Mode 4 and IEEE 802.11p. However, IEEE 802.11p can be as good as Mode 4 in terms of PDR and better in terms of UD up to a certain distance by a suitable choice of MCS. In contrast to [73], the two papers [75] and [74] do not consider additional occupancy of the cellular network by non-V2X applications.

[76] investigates LTE-V2X Mode 3 and 4 and IEEE 802.11p with respect to their suitability for truck platooning in a high-traffic scenario. The results show that LTE-V2X Mode 3 performs better than 802.11p in terms of reliability for this use case, and LTE-V2X Mode 4 performs only marginally better than 802.11p. One possible explanation is provided by collision errors: IEEE 802.11p is more prone to collisions due to its contention-based mechanism. LTE-V2X Mode-4 also faces collisions but to a lesser extent. Mode-3 effectively avoids collisions through centralized scheduling. Thus, the authors consider LTE-V2X to be the better technology for truck platooning and high-density scenarios in general.

In [77], different MCSs for LTE-V2X and IEEE 802.11p were tried and compared. The authors were able to show that the use of different MCSs can improve the performance of IEEE 802.11p especially for long ranges. For IEEE 802.11p, the bit rate was improved to a range from 3 to 27 Mb/s. In contrast, the bit rate in LTE-V2V ranges between 1.15 Mb/s and. 17.71 Mb/s. However, the authors still conclude that, while other metrics being equal, LTE-V2V offers a higher range and is better suited when a high perceptual range is the goal.

In [49], the authors R. Sattiraju et al. compare LTE-V2X and IEEE 802.11p with respect to their link level performance of the physical layer (PHY) in different transmission channels. As a metric, they use Block Error Rate (BLER) vs. Signal-to-Noise Ratio (SNR). The BLER is the ratio between the number of blocks transmitted with errors and the total number of blocks transmitted. Their results show that LTE-V2X offers significantly better performance in the physical layer than IEEE 802.11p, and the authors also emphasize that LTE-V2X is better suited for high vehicle speeds. The effective range is inferred from the performance over different SNR levels and channel conditions. LTE-V2X, with its better channel coding and estimation, is likely to maintain reliable communication over longer distances compared to ITS-G5.

V. Mannoni et al. [42] compare LTE-V2X PC5 Mode 4 and ITS-G5 at both PHY and MAC levels. They find LTE-V2X offers higher flexibility and better SNR performance at the same data rate. However, LTE-V2X degrades faster with increasing user density, making ITS-G5 preferable at densities >150 users/km². ITS-G5 exhibits lower latency over short distances, but LTE-V2X may have smaller latency at longer distances under certain settings. The average bit rate for ITS-G5 was evaluated to be approximately 4.76 Mb/s using QPSK modulation a coding rate of ½. C-V2X on the other hand supports variable bit rates. For instance, a configuration with 48 Resource Blocks (RBs) and MCS 3 offers 2.4 Mb/s. Other configurations can provide throughputs from 1.6 Mb/s to 6.4 Mb/s depending on the message size and the number of RBs.

In [78], R. Roux et al. analyze the physical layer performance of ITS-G5 and LTE-V2X PC5 Mode 4. The paper compares ITS-G5 and LTE-V2X (PC5 mode 4) technologies focusing on Packet Delivery Ratio (PDR), latency, and range. LTE-V2X shows superior PDR performance, achieving a range of 35 meters at a PDR of 0.9, compared to 23 meters for ITS-G5. The study highlights that LTE-V2X minimizes latency by design, whereas ITS-G5 can introduce significant latency under high traffic conditions. Both technologies are evaluated for Channel Busy Ratio (CBR) under high traffic, with configurations ensuring efficient congestion control. The results indicate that LTE-V2X outperforms ITS-G5 in terms of packet delivery and range, demonstrating better performance in handling propagation and collision errors.

The authors in [38] analyze the performance of V2X approaches regarding Ultra Reliable Low Latency

Communications and enhanced Mobile Broadband applications in simulation. NR-V2X outperforms all other standards in terms of reliability, range, latency, and data rates, making it the most suitable for advanced V2V communications. IEEE 802.11bd shows significant improvements over IEEE 802.11p with better packet error rates (PER), increased range, and higher throughput. LTE-V2X, while better than IEEE 802.11p in data rates and reliability, is surpassed by NR-V2X and IEEE 802.11bd in overall performance, particularly in latency and packet reception ratio (PRR). Overall, NR-V2X and IEEE 802.11bd emerge as the leading technologies, with NR-V2X having a slight edge in data rates and latency.

In [79], Zugno et al. analyze the end-to-end performance of NR-V2X communications at millimeter waves as a simulation with two vehicles. In that context, the authors investigate the network performance considering direct V2V communication and the impact on antenna array, communication distance, modulation coding scheme and others. Their results show a small latency by using mmWave with a modulation coding scheme MCS 28 in short communication distances because of bad channel states. For small size packets as tested, MCS 0 results in a low latency for communication distances up to 500 m. Packet loss rate is analyzed through PDR, with higher losses observed at increased distances, especially in NLOS conditions. Average latency is significantly affected by numerology and MCS, with numerology 3 generally providing lower delays at shorter distances, though latency increases with distance and NLOS conditions.

Todisco et al. [57] are focusing on the performance evaluation of the NR-V2X Mode 2 SL using an open-source simulator. Thereby the authors show that blind retransmission (HARQ) is only useful at low traffic-density. In high traffic-density scenarios, HARQ can further increase the network load. In addition, they showed the influence of the modulation coding scheme on the possible package size and transmission distance at different traffic densities using a system-level simulator. With an increasing packet size, the sub carrier spacing needs to be reduced, while the MCS needs to be increased for more resources per time.

Saad et al. compare the resource selection performance of NR-V2X Mode 2 with LTE-V2X Mode 4 in [80] using ns-3 simulator framework. For their analysis, the authors use Cooperative Awareness Massages (CAM) with variable size according to the ETSI specification. The results show a slightly decreasing packet delivery ration (PDR) for NR-V2X with an increasing number of vehicles from 100 to 300 at subcarrier spacings of 30 kHz and 60 kHz. For LTE-V2X the PDR drops significantly from 70 % to 50 %. Furthermore, it can be seen in the results that with an increasing number of Sidelink subchannels and a longer Resource Reservation Time, a higher PDR can be achieved for NR-V2X and LTE-V2X. Nevertheless, NR-V2X Mode 2 outperforms LTE-V2X Mode 4 in each of the executed tests.

[81] concentrates on the comparison of 5G NR-V2X communication with LTE-V2X communication in the context of the vehicle platooning use case. They present an analytical model for evaluating the performance which considers aspects such as multiple access interference in vehicle platooning scenarios and places emphasis on resource allocation in LTE-V2X, which significantly influences the performance of Sidelink communication in real-world deployment scenarios. The comparative study confirms the superior performance of NR-V2X over LTE-V2X, especially in scenarios with a large inter-vehicle distance and a high number of vehicles. The CBR is indirectly evaluated through the effectiveness of the sensing-based semi-persistent scheduling (SB-SPS) scheme in allocating resources and minimizing interference. NR-V2X's improved sensing mechanisms lead to better resource allocation and lower channel busy ratio compared to LTE-V2X.

The impact on co-channel coexistence between ITS-G5 and LTE-V2X mode 4 is analyzed by Roux et al. in [82]. For the evaluation of the influence of the coexistence, the authors consider only an urban scenario in a ns3-based simulation with one 10 MHz channel shared between ITS-G5 and LTE-V2X Mode 4. Their results show that with an equal number of 800 ITS-G5 and LTE-V2X equipped vehicles, the ITS-G5 communication benefits from applying the coexistence method. LTE-V2X Mode 4 shows a degradation at the same time, as resources for the resource selection is reserved for ITS-G5. CBR is discussed as well, noting that ITS-G5 has an average channel occupancy duration of 529 μs per CAM message, whereas LTE-V2X (using MCS3) has an occupancy of 1500 μs per message. This higher occupancy time in LTE-V2X results in more frequent collisions.

*C. Technology assessment*

Table 4 and Table 5 provide a condensed overview of the performance comparisons between different communication approaches. To summarize, this paper expands on [44] by adding a comparison between NR-V2X and other approaches as well as additional sources overall. While no conclusion can be drawn for the comparison between LTE-V2X Mode 3 and NR-V2X due to the lack of related work, it appears as if NR-V2X outperforms LTE-V2X and the IEEE standards in all low and most of the high traffic scenarios. However, due to different approaches in conducting the measurements across papers, e.g., by measuring different KPIs, it remains challenging to create a clear overall picture. Furthermore, while the general methodology of using traffic simulations for data acquisition is beneficial in terms of comparability between IEEE 802.11 and the other approaches, no conclusion can be drawn for the performance in specific V2X use cases or real world V2X scenarios. As such, the authors follow that there still exists a research gap in networking performance evaluation that is yet to be explored.

Table 4: Literature overview about IEEE802.11 based approaches compared to Vehicle-to-Everything approaches.

| Ref | Method | Compared to | Result |
|---|---|---|---|
| [83–88] | IEEE 802.11p | - | As vehicle speed and communication distance increase, packet delivery ratio decreases while transmission latency remains constant. In addition, vehicle density only affects packet delivery ratio for a given packet size in networks that are not fully connected. |
| [38, 41, 89] | | IEEE 802.11 bd | IEEE 802.11bd consistently outperforms 802.11p in all scenarios for communication distance, latency, throughput, and packet delivery ratio for both small and large packet sizes. |
| [38, 49, 73, 90] | | LTE-V2X | LTE-V2X outperforms IEEE 802.11p in all scenarios for communication distance, throughput, and packet delivery ratio for both small and large packet sizes. For small packet sizes, IEEE 802.11p has lower latency for short- and medium-range communications. For larger packets, LTE-V2X outperforms IEEE 802.11p in terms of latency. |
| [2, 75, 77, 91] | | LTE-V2X Mode 3 | IEEE 802.11p is robust at limited distances while LTE-V2X is more reliable in mid- and long-range situation. The packet loss and the latency of IEEE 802.11p is lower than the latency of LTE-V2X in PC5 and Uu communication. |
| [42–44, 51, 58, 72, 75, 76, 78, 82, 91, 92] | | LTE-V2X Mode 4 | IEEE 802.11p has a lower latency than LTE-V2X, while the PDR in LTE-V2X is higher compared to IEEE 802.11p in V2I scenarios. When vehicles are traveling in the opposite direction, the packet delivery ratio is higher and the latency is lower with IEEE802.11p. PDR for both approaches is affected by the number of vehicles and their speed in the communication area. |
| [38, 93] | | NR-V2X | NR-V2X outperforms IEEE 802.11p in all scenarios for communication distance, latency, throughput, and packet delivery ratio for both small and large packet sizes. |
| | | NR-V2X Mode 1 | Not available in literature |
| [94] | | NR-V2X Mode 2 | NR-V2X Mode 2 outperforms IEEE 802.11p in terms of packet delivery ratio and communication distance. While vehicle density reduces the PDR over the communication distance, NR-V2X Mode 2 has little effect on the PDR. For both technologies, the PDR decreases as the number of packets transmitted increases. NR-V2X Mode 2 is more resilient to an increase in the number of packets transmitted. |
| | IEEE 802.11bd | - | Not available in literature |
| [38, 93] | | LTE-V2X | LTE-V2X outperforms IEEE 802.11bd in PDR. In small packet size scenarios, LTE-V2X shows higher throughput. At the same time, the latency of IEEE 802.11bd is lower in short- and medium-range scenarios. In larger packet size scenarios, 802.11bd shows higher throughput and lower latency in near field communication. |
| | | LTE-V2X Mode 3 | Not available in literature |
| | | LTE-V2X Mode 4 | Not available in literature |
| [38, 93, 95] | | NR-V2X | NR-V2X outperforms IEEE 802.11bd in PDR. In small packet size scenarios, NR-V2X shows higher throughput. At the same time, the latency of IEEE 802.11bd is lower in short- and medium-range scenarios. In larger packet size scenarios, 802.11bd shows higher throughput and lower latency in short-range communications. |
| | | NR-V2X Mode 1 | Not available in literature |
| [39] | | NR-V2X Mode 2 | No clear statement on performance. However, since NR-V2X does not require backward compatibility, it can use the latest and most advanced MAC and PHY layer techniques. |

Table 5: Literature overview about cellular-based approaches compared to Vehicle-to-Everything approaches.

| Ref | Method | Compared to | Result |
|---|---|---|---|
| [83, 96] | LTE-V2X | - | Latency and Packet Delivery Ratio depend on relative vehicle velocity. Latency increases with higher speeds, while PDR decreases. Sidelink communication reduces latency and increases PDR. |
| [38, 93] | | NR-V2X | NR-V2X outperforms LTE-V2X in all scenarios in terms of communication distance, throughput and packet delivery ratio for both small and large packet sizes. LTE-V2X has lower latency than NR-V2X in long-distance communication scenarios with small packet sizes. |
| | LTE-V2X Mode 3 | - | Not available in literature |
| [37, 75, 91] | | LTE-V2X Mode 4 | The Packet Delivery Ratio in LTE-V2X Mode 3 and Mode 4 depends on the number of packets sent per second and the vehicle density. An increasing number of packets per second and a higher vehicle density show that Mode 3 can maintain a high packet delivery ratio over the communication distance. Mode 3 outperforms Mode 4 in terms of PDR in all situations such as highway, rural or urban scenarios. |
| | | NR-V2X Mode 1 | Not available in literature |
| | | NR-V2X Mode 2 | Not available in literature |
| [56, 69, 97–99] | LTE-V2X Mode 4 | - | Packet Delivery Ratio is affected by the Modulation Coding Scheme (MCS) and the number of vehicles in the communication range, resource blocks and message size. Lower MCS results in higher PDR in long range situations. In high density situations, the Packet Delivery Ratio decreases due to packet collision. |
| [1, 81] | | NR-V2X | The Packet Delivery Ratio for LTE-V2X Mode 4 and NR-V2X is affected by the modulation coding scheme, the vehicle density and the number of transmitted messages. NR-V2X outperforms LTE-V2X Mode 4 in every situation. |
| | | NR-V2X Mode 1 | Not available in literature |
| [80, 81, 94] | | NR-V2X Mode 2 | The Packet Delivery Ratio of both NR-V2X Mode 2 and LTE-V2X Mode 4 depends on the vehicle velocity and the number of transmitted packets. However, NR-V2X Mode 2 outperforms LTE-V2X Mode 4 because the influence of velocity and transmission rate is lower. |
| [79, 100, 101] | NR V2X | | For a Modulation Coding Scheme (MCS) less than 27, vehicle velocity has no effect on transmission efficiency. For an MCS greater than or equal to 27, performance degrades as velocity increases. In addition, a higher MCS improves throughput. For an MCS greater than or equal to 27, throughput is dependent on vehicle velocity. |
| | NR V2X Mode 1 | - | Not available in literature |
| | | NR V2X Mode 2 | Not available in literature |
| [57, 91, 102] | NR V2X Mode 2 | - | The selection of packet generation rules from ETSI and 3GPP affects the Packet Delivery Ratio. Using the ETSI rules, the PDR is higher compared to using the 3GPP generation rules. NR-V2X Mode 2 is affected by the numerology, the number of PSSCH transmissions, the resource selection window, the modulation coding scheme, and the maximum number of resources per reservation. |

## VI. GEOGRAPHICAL USAGE OF V2X

This section introduces the worldwide distribution of IEEE 802.11 and C-V2X. The following sections are divided into regions because some regularities are defined across countries. In particular, the geographical regions of the United States of America, Europe, and China are considered.

*A. United States of Amerika*

In the United States, DSRC was introduced in 1999 [44]. It was the dominant technology until 2020, which slowed down both laboratory and field testing of other communication standards. During this period, several DSRC field tests were conducted, such as the SPaT Challenge [103]. In SPaT, traffic light cycle time was distributed using DSRC. According to the US Department of Transportation (USDoT), more than 70 DSRC applications were active in the US in 2018. [104, 105]. Due to the increasing importance of C-V2X and the low market penetration of DSRC, the Federal Communications Commission (FCC), which regulates communications frequencies, has decided to reuse parts of the 5.9 GHz spectrum for other V2X-related applications in 2019 [106] and issued an order in November 2020 [107]. The frequency range between 5850 MHz and 5895MHz is re-allocated for the expansion of midband spectrum operations, while the frequency range between 5895 MHz and 5925 MHz is now used for C-V2X communication. In 2021, Audi started two C-V2X projects in the 5.9 GHz range [108] using LTE radio and the PC5 interface. According to market analysts, even Toyota, a former advocate of DSRC in 2016 [19] changed his plans regarding the usage of C-V2X [109]. It is therefore reasonable to assume that C-V2X will replace DSRC in the long term [110]. Based on the current changes, it can be expected, that the market penetration of C-V2X will increase until 2025. It can further be assumed that the first commercial products equipped with NR-V2X will be developed. As the publication of the next IEEE 802.11bd standard is expected in 2030, it is likely that researchers and companies also try to establish new DSRC use cases till then.

*B. Europe*

Instead of using DSRC, ITS-G5 technology based on IEEE 802.11p was introduced and tested in Europe. A first proposal from the European Union to use ITS-G5 as the main technology was rejected in 2019 [111], where it was argued that defining a core technology would stifle technological progress. Subsequently, the ETSI published a standard defining the C-V2X technology as the access layer for intelligent transportation systems. In this standard, C-V2X communication over the PC5 interface uses the same frequencies as the ITS-G5 technology [112]. For the sake of completeness, it should be mentioned that the coexistence of ITS-G5 and C-V2X is possible. ETSI has published two technical reports in 2021 that address the coexistence of ITS-G5 and C-V2X SL using the PC5 interface [113].

However, not all car manufacturers rely on ITS-G5. Market research companies consider C-V2X technology to be superior, despite its possible coexistence with ITS-G5 [114]. As of 2020, Volkswagen vehicles such as the VW Golf MK8 and VW ID models support ITS-G5. Frost & Sullivan [109] concludes that Volkswagen is likely to move to NR-V2X at a later stage when it is well developed and researched. This assumption can be supported by an analysis of recently started projects: As of 2021 [110], there are seven projects testing the viability and performance of C-V2X in Europe. Thus, the trend towards C-V2X is also visible.

The mass deployment of ITS-G5 and C-V2X was planned for 2025, but according to recently published information to Euro NCAP (2017) [115, 116], it is delayed to 2027. Nevertheless, it is expected that Europe's technology openness will lead to an almost equal distribution of communication systems using ITS-G5 and C-V2X until 2027 and NR-V2X beyond 2030 due to the different interests of nations within the European Union. The technology openness allows flexible adaptation to use cases and optimization of bandwidth requirements and transmission distance.

*C. China*

China has never deployed DSRC. Instead, they have directly focused on C-V2X. The first test of C-V2X was conducted in 2016 [117]. This was followed by interoperability of LTE-V2X devices from different manufacturers using the PC5 interface by 2018. Mass production of vehicles with V2X functionality began in 2020 [110]. In addition, C-V2X is considered a core technology to enable the development of a national traffic control network connecting roads and vehicles using 5G communication [114]. Therefore, China enacted a national strategy for the Internet of Vehicles (IoV) in 2021, which includes V2X communication [110]. The 20 MHz frequency band of C-V2X is defined from 5905 MHz to 5925 MHz. In general, China is already well advanced in terms of 5G network deployment. According to Businesswire [118], by the end of 2021, all major cities and at least 97% of counties were covered by 5G. The penetration of NR-V2X-enabled vehicles in 2021 was less than 0.4% of the total number of vehicles with communication capability but is expected to grow to 15% by 2025.

As the first NR-V2X equipped vehicles are already available in China, the fast rise of the technology is highly likely. Based on plans as of 2022, there shall be a market penetration of up to 50 % for C-V2X in 2030. Furthermore, it can be assumed that NR-V2X will be the mainly focused technology for vehicle communication in 2030 and beyond, as the first commercial applications shall be introduced in 2025 [118, 119].

Some of the authors conclude that bandwidth and latency requirements will become more relevant as new use cases for connected and autonomous driving, such as collective perception or extended sensors, are deployed, leading to increased market penetration of NR-V2X by 2030 and beyond

## VII. CHALLENGES

V2X communication is a key technology to increase road safety and traffic efficiency [3, 72]. However, there are still some open challenges pending regarding the overall vehicle communication, including the required data rate and bandwidth as well as the message generation rules. The following sections describe these challenges in detail to clarify some open research areas and specific open topics from a use case specific perspective.

To achieve the goals of autonomous driving, a detailed perception of the environment is required [66]. This can be realized by data and information exchange [120]. But due to dynamic influences like lightning, weather and occlusion to participants within the environment, discrepancies in perception can occur [121], thus making it difficult to maintain the *integrity of perceived environment* (section VII.A). Additionally, as several participants act in the same environment at the same time, they will detect the same objects and share them with others [122, 123]. This increases the network load but does not necessarily provide any additional benefit. The rate of redundant information will increase with the number of participants. Thus, *managing data transmission in dynamic environments* (section VII.B) is crucial. Especially in the context of V2X, various use cases (see section IV.D) that require the exchange of data and information between involved participants on different levels exist. Thus, the system efficiency and performance can be directly influenced by the *communication architecture and function offloading* (section VII.C).

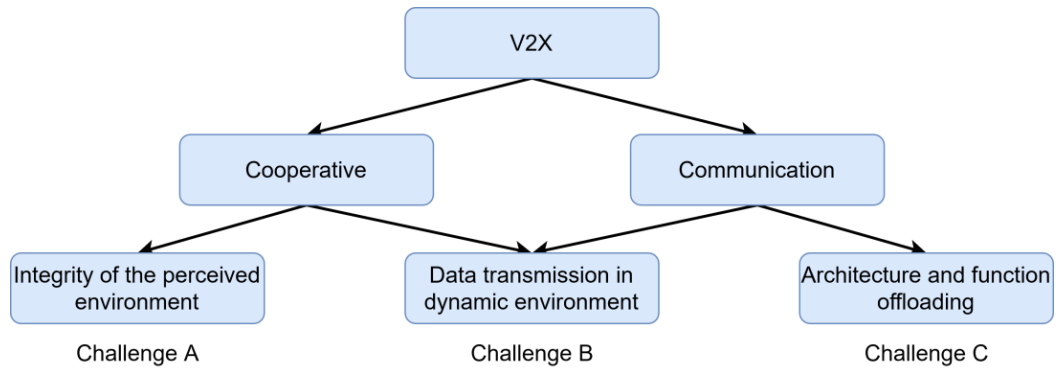

Figure 5: Overview about open challenges in the field of V2X communication.

Figure 5 summarizes the above-mentioned points and clarifies the considered topics in different sections. The following subsections first derive the challenges (Cx.y) in the context of the specified topic and finally provide a summarized overview of open challenges.

A. *Challenge: Integrity of the perceived environment*

To improve the efficiency and safety in the mobility sector, critical situations and ego-vehicles' information needs to be transmitted between nearby participants. In addition, knowledge about all objects within a defined area in the vehicles' environment is required. The currently established V2X use cases try to fulfill these requirements through specified message formats that enable the transmission of required information [124–127]. In this context, the vehicle status information or detected objects within the ego vehicle's environment are transmitted. A vehicle equipped with sensors and the capability to detect its environment is called ego vehicle.

To save bandwidth, aggregated information such as detected objects or identified situations are transmitted [126], in e.g. cooperative perception, 35 bytes are required to transmit an object in a collective perception message. An object needs to be identified before its transmission, so even with e.g. cooperative perception, a driver or a vehicle can never assume that all present objects within a FoV are known [121]. Exchanging identified objects within a vehicle's FoV can close blind spots caused by occlusion, but cannot increase detection confidence [121, 128] (CA.1). To raise the efficiency and the confidence in the environmental model, new approaches to detect the surroundings in more detail are required. Possible ideas are the transmission of sensor raw data [129] and the transmission of feature maps [130]. While the transmission of sensor data results in a more detailed representation of the environment, as all perception data is available, it requires high bandwidth for data transmission at the same time [131]. As the estimated data generated by an autonomous vehicle is about 4 TB [132] per day and depending on the sensor set, between 50 MB and 70 MB per second [133], this approach seems to be difficult without any compression strategy, data pre-processing or intelligent data selection. The application of compression strategies [134] or the intelligent selection of data can be helpful to enable use cases requiring cooperative interaction (see Section IV.D), but still results in higher bandwidth usage (CA.2). Besides classical compression, a more efficient approach is to transmit the feature maps generated by machine learning algorithms in connected vehicles [130], but still fail in large environments (CA.3) due to alignment errors and hardware and storage requirements. Since the transmitted maps needs to be aligned and fused, the temporal and spatial domain needs to be considered [135]. Since the interaction time between several road users is short, the temporal aspect cannot be neglected, and high computational power is required to solve the task during the interaction time (CA.4). To enable cooperative use cases, information from different vehicles is to be fused. However, since sensors and detection algorithms can cause errors due to distance to an object, occlusion, weather condition and others, inconsistent information will occur, which needs to be resolved (CA.5).

In conclusion, the following five main challenges are identified regarding the integrity of the perceived environment:
- (CA.1) cooperative information exchange is limited by only considering identified objects,
- (CA.2) the transmission of unprocessed sensor data and feature maps results in a higher network load,
- (CA.3) cooperative SLAM and feature maps are not well appropriate for large scale maps,
- (CA.4) short interaction time between participants,
- (CA.5) inconsistent information needs to be resolved.

B. *Challenge: Managing data transmission in dynamic environments*

Since the transmission of data and information is crucial for automated and autonomous driving and it is assumed that an autonomous vehicle Is expected to generate up to 4 TB of data a day, the network resources have to be used as efficiently as possible. As there are multiple vehicles perceiving a situation (and thus, the objects involved) in the same environment, data redundancy is to be expected according to [122, 136, 137]. Especially in medium- and high-density traffic scenarios, it is likely that information regarding specific objects will be transmitted and received several times (CB.1) [138], which dissipates rare network resources. This can lead to network congestions, resulting in higher latency and thus delayed message transmission. This behavior is not limited to specific messages, but to all cyclically transmitted message packets. Based on that, the Decentralized Congestion Control (DCC) queues messages in overload situations using a First In First Out (FIFO) buffer [139]. According to Delooz et al. [123], this mechanism can reduce network latency by reducing the number of messages sent. However, relevant messages have to wait in a queue for an undefined period of time, resulting in a delayed transmission of information and, since the interaction time between clients is short, invalid messages. Hence, the ETSI specification defines the usage of four message generation rules for Cooperative Perception Messages to avoid network conjunction [124]. Due to the low market penetration of V2X-equipped vehicles and roadside units, the verification of the functions and the effects between them have only been tested in simulations (CB.2).

Based on this review, it is shown that the different V2X approaches (see Sections IV.A and IV.B) try to solve the same problem by using different technologies and, most importantly, individually designed frequency channels, sometimes overlapping with bands of other V2X approaches and protocol stacks. In a globalized world, it is likely that vehicles produced for the U.S., European, Chinese or any other market will be driven on other continents, which would ultimately lead to a loss in functionality. This is the case even though DSRC and ITS-G5 are based on IEEE 802.11, since they differ in detail like the channel's frequency bands and different protocol stacks. In addition, the C-V2X Uu interface uses the same frequency as the ITS-G5 (CB.3), which may result in a loss of performance [82].

In conclusion, the following three main challenges are identified:
- (CB.1) reduce rate of information redundancy in mid- and high-density traffic scenarios,
- (CB.2) Field tests in real environments are difficult,
- (CB.3) Continent specific and thus incompatible V2X paradigms.

C. *Challenge: Communication architecture and function offloading*

Based on the introduced use cases in section IV.D and the available network components such as cloud, fog, edge and vehicle communication [140], various possibilities exist to implement a performant communication architecture. While some functions like cooperative mapping benefit from the centralized provision of information, functions such as remote driving require the instantaneous exchange of information. All use cases have with specific requirements in terms of latency, bandwidth, computational capacity and other factors [141, 142]. Thus, it is crucial to elaborate which use cases can be applied to the different architectures and how this affects their use case performance parameters (CC.1). The dynamic behavior of vehicles results in short interaction times between different participants, making it even more difficult to provide accurate detection results and decision making through cooperative functions.

The corresponding morphological box is shown in Table 6. The morphological combination of five parameters results in a complex solution space that enables the determination of possible communication architectures through permutation. The *Main Component* defines the type of nodes available in the network, while *V2X technology* represents the approach itself. Considering only these two parameters, four architectures with four different performances are possible. Additional parameters such as *the number of participants* and the *communication distance* or the *transmitted data* itself will directly influence the performance of the system, and thus resulting in a huge variance of use case specific architectures (CC.2).

Due to the progress in assisted and autonomous driving and with respect to cooperative vehicle functions (CVF), the complexity of vehicular functions is continuously increasing [143]. Thus, additional computational and storage resources are needed to meet the requirements regarding processing time and accuracy [144]. As modern vehicle functions are steadily evolving, increasing computational and storage requirements are expected [143]. If these requirements cannot be fulfilled by a vehicle, the latest updates or the whole functionality will not be available for further usage [145]. CVF especially require data exchange between all involved participants [138]. If the CVF is deployed and executed locally, the same message needs to be transmitted between each participating vehicle. Thus, increasing the network load and the information redundancy in the network. In this context, function offloading describes the possibility to deploy vehicle functions to network elements such as the edge, fog, or cloud, and can therefore mitigate these initial issues [143], but may result in higher functional latency [146]. With an increasing number of vehicle functions and their complexity, it is hard to decide whether a vehicle function can be offloaded to the edge, fog, or cloud, or whether it needs to be deployed in a vehicle (CC.3). Hence, each functionality and use case requires a specific maximum latency and minimum bandwidth to maintain its functionality and a detailed knowledge of these thresholds is needed (CC.4). To gain the latter, extensive performance evaluations of networks for specific use cases in real-world networks using real clients while considering different Quality of Service parameters (QoS) are required (CC.5). Summarized, the following five major challenges have been identified with respect to possible communication architectures and function offloading:

Table 6: Possible parameters influencing use case performance

| Parameter | Characteristics | | | |
|---|---|---|---|---|
| Main component | Cloud | Edge | Fog | Vehicles |
| V2X technology | DSRC | ITS-G5 | LTE-V2X | NR-V2X |
| Number of participants | 1 | 10 | 100 | 1000 |
| Communication distance | <50 m | <100 m | <150 m | <200 m |
| Transmitted data | objects | data | Compressed data or features | hybrid |

- (CC.1) impact of dynamic network parameters on dynamic use cases is unknown,
- (CC.2) huge variety on use case specific network topologies,
- (CC.3) decision, what function can be offloaded to the network backend,
- (CC.4) detailed knowledge about performance thresholds is required,
- (CC.5) use case specific evaluation of network performance parameters in the real world necessary.

## VIII. CONCLUSION

In this paper, the most popular Vehicle-to-Everything communication standards based on IEEE 802.11 (DSRC and ITS-G5) and 4th and 5th generation cellular standards (C-V2X, LTE-V2X, NR-V2X) have been analyzed in terms of their performance, current and future geographical usage, and current challenges. The main findings can be summarized as follows:

1. The analysis of the existing literature shows that NR-V2X outperforms IEEE 802.11p and C-V2X respectively LTE-V2X in all scenarios (low- and high-density traffic). Compared to NR-V2X, IEEE 802.11bd shows higher performance (latency and throughput) in short-range communication as the packet size increases.
2. A gradual shift from DSRC and ITS-G5 to C-V2X by 2030 and NR-V2X after 2030 is expected globally.
3. We have identified three main challenges based on our literature research:
   a. maintaining the integrity of the perceived environment: Highly dynamic environment with short interaction time and heterogeneous clients perceiving the same environment from different perspectives.
   b. managing cooperative data transmission in dynamic environments: Optimal decision making requires large amounts of data, but because of a), data redundancy, contradictions and incompatible protocols must be considered.
   c. possible communication architectures and function offloading: The execution of cooperative tasks requires a stable network connection with respect to specific quality of service requirements in real-world environments.

Our analysis shows that NR-V2X currently has the highest performance. However, it is also clear that IEEE 802.11bd can provide an advantage in certain situations, but this is not yet of great importance due to the late expected release period. Also due to the increasing degree of autonomy of vehicles, it is expected that the focus in the near future will be on the use of NR-V2X.

To enable the use of V2X in different scenarios, function offloading and data flow optimization are crucial. Future research needs to explore how vehicle functions can be deployed in distributed cloud/edge systems and how the associated data streams can be efficiently orchestrated and managed. In this context, how to reduce the operational complexity of such systems through continuous analysis and optimization needs to be considered.

## REFERENCES

[1] C. Campolo, A. Molinaro, F. Romeo, A. Bazzi, and A. O. Berthet, "5G NR V2X: On the Impact of a Flexible Numerology on the Autonomous Sidelink Mode," in *2019 IEEE 2nd 5G World Forum (5GWF)*, Dresden, Germany, 2019, pp. 102–107, doi: 10.1109/5GWF.2019.8911694.

[2] Aman Kumar Gulia, "A Simulation Study on the Performance Comparison of the V2X Communication Systems: ITSG5 and C-V2X Communication Systems: ITS-G5 and C-V2X," Masterthesis, The School of Electrical Engineering and Computer Science, KTH Royal Institute of Technology, Stockholm, Sweden, 2020.

[3] H. Zhou, W. Xu, J. Chen, and W. Wang, "Evolutionary V2X Technologies Toward the Internet of Vehicles: Challenges and Opportunities," *Proc. IEEE*, vol. 108, no. 2, pp. 308–323, 2020, doi: 10.1109/JPROC.2019.2961937.

[4] S. Chen, J. Hu, Y. Shi, L. Zhao, and W. Li, "A Vision of C-V2X: Technologies, Field Testing, and Challenges With Chinese Development," *IEEE Internet Things J.*, vol. 7, no. 5, pp. 3872–3881, 2020, doi: 10.1109/JIOT.2020.2974823.

[5] M. H. C. Garcia *et al.*, "A Tutorial on 5G NR V2X Communications," *IEEE Commun. Surv. Tutorials*, vol. 23, no. 3, pp. 1972–2026, 2021, doi: 10.1109/COMST.2021.3057017.

[6] Falk Dettinger, Matthias Weiß, and Michael Weyrich, Eds., *Future Use Cases for Vehicular Communication based on Connected Functions*. 2024 IEEE 100th Vehicular Technology Conference (VTC2024-Fall): IEEE, 2024, doi: print.

[7] R. Sedar, C. Kalalas, F. Vazquez-Gallego, L. Alonso, and J. Alonso-Zarate, "A Comprehensive Survey of V2X Cybersecurity Mechanisms and Future Research Paths," *IEEE Open J. Commun. Soc.*, vol. 4, pp. 325–391, 2023, doi: 10.1109/OJCOMS.2023.3239115.

[8] A. Alnasser, H. Sun, and J. Jiang, "Cyber security challenges and solutions for V2X communications: A survey," *COMPUTER NETWORKS*, vol. 151, pp. 52–67, 2019, doi: 10.1016/j.comnet.2018.12.018.

[9] M. N.-E. Saulaiman, M. Kozlovszky, and A. Csilling, "A Survey on Vulnerabilities and Classification of Cyber-Attacks on 5G-V2X," in *2021 IEEE 21st International Symposium on Computational Intelligence and Informatics (CINTI)*, Budapest, Hungary, 2021, pp. 235–240, doi: 10.1109/cinti53070.2021.9668440.

[10] T. Yoshizawa *et al.*, "A Survey of Security and Privacy Issues in V2X Communication Systems," *ACM Comput. Surv.*, vol. 55, no. 9, pp. 1–36, 2023, doi: 10.1145/3558052.

[11] J. Huang, D. Fang, Y. Qian, and R. Q. Hu, "Recent Advances and Challenges in Security and Privacy for V2X Communications," *IEEE Open J. Veh. Technol.*, vol. 1, pp. 244–266, 2020, doi: 10.1109/OJVT.2020.2999885.

[12] F. A. Schiegg, I. Llatser, D. Bischoff, and G. Volk, "Collective Perception: A Safety Perspective," *Sensors*, early access. doi: 10.3390/s21010159.

[13] I. Llatser, T. Michalke, M. Dolgov, F. Wildschutte, and H. Fuchs, "Cooperative Automated Driving Use Cases for 5G V2X Communication," in 2019, doi: 10.1109/5GWF.2019.8911628.

[14] L. Hobert, A. Festag, I. Llatser, L. Altomare, F. Visintainer, and A. Kovacs, "Enhancements of V2X communication in support of cooperative autonomous driving," *IEEE Commun. Mag.*, vol. 53, no. 12, pp. 64–70, 2015, doi: 10.1109/MCOM.2015.7355568.

[15] F. A. Schiegg, N. Brahmi, and I. Llatser, "Analytical Performance Evaluation of the Collective Perception Service in C-V2X Mode 4 Networks," in *2019 IEEE Intelligent Transportation Systems Conference (ITSC)*, Auckland, New Zealand, uuuu-uuuu, pp. 181–188, doi: 10.1109/ITSC.2019.8917214.

[16] C. R. Storck and F. Duarte-Figueiredo, "A Survey of 5G Technology Evolution, Standards, and Infrastructure Associated With Vehicle-to-Everything Communications by Internet of Vehicles," *IEEE Access*,

vol. 8, pp. 117593–117614, 2020, doi: 10.1109/ACCESS.2020.3004779.

[17] H. Abou-zeid, F. Pervez, A. Adinoyi, M. Aljlayl, and H. Yanikomeroglu, "Cellular V2X Transmission for Connected and Autonomous Vehicles Standardization, Applications, and Enabling Technologies," *IEEE Consumer Electron. Mag.*, vol. 8, no. 6, pp. 91–98, 2019, doi: 10.1109/MCE.2019.2941467.

[18] K. Garlichs, H. Günther, and L. C. Wolf, "Generation Rules for the Collective Perception Service," in *2019 IEEE Vehicular Networking Conference (VNC)*, 2019, pp. 1–8, doi: 10.1109/VNC48660.2019.9062827.

[19] M. Boban, A. Kousaridas, K. Manolakis, J. Eichinger, and W. Xu, "Connected Roads of the Future: Use Cases, Requirements, and Design Considerations for Vehicle-to-Everything Communications," *IEEE Vehicular Technology Magazine*, vol. 13, no. 3, pp. 110–123, 2018, doi: 10.1109/MVT.2017.2777259.

[20] S. Sun, J. Hu, Y. Peng, X. Pan, L. Zhao, and J. Fang, "Support for vehicle-to-everything services based on LTE," *IEEE Wireless Commun.*, vol. 23, no. 3, pp. 4–8, 2016, doi: 10.1109/MWC.2016.7498068.

[21] V. V. Chetlur and H. S. Dhillon, "Coverage and Rate Analysis of Downlink Cellular Vehicle-to-Everything (C-V2X) Communication," *IEEE Trans. Wireless Commun.*, vol. 19, no. 3, pp. 1738–1753, 2020, doi: 10.1109/TWC.2019.2957222.

[22] E. Uhlemann, "Initial Steps Toward a Cellular Vehicle-to-Everything Standard [Connected Vehicles]," *IEEE Vehicular Technology Magazine*, vol. 12, no. 1, pp. 14–19, 2017, doi: 10.1109/MVT.2016.2641139.

[23] N. Raza, S. Jabbar, J. Han, and K. Han, "Social vehicle-to-everything (V2X) communication model for intelligent transportation systems based on 5G scenario," in *Proceedings of the 2nd International Conference on Future Networks and Distributed Systems*, Amman Jordan, A. Abuarqoub, B. Adebisi, M. Hammoudeh, S. Murad, and M. Arioua, Eds., 2018, pp. 1–8, doi: 10.1145/3231053.3231120.

[24] M. Hasan, S. Mohan, T. Shimizu, and H. Lu, "Securing Vehicle-to-Everything (V2X) Communication Platforms," *IEEE Trans. Intell. Veh.*, vol. 5, no. 4, pp. 693–713, 2020, doi: 10.1109/TIV.2020.2987430.

[25] A. Alalewi, I. Dayoub, and S. Cherkaoui, "On 5G-V2X Use Cases and Enabling Technologies: A Comprehensive Survey," *IEEE Access*, vol. 9, pp. 107710–107737, 2021, doi: 10.1109/ACCESS.2021.3100472.

[26] R. Khezri, D. Steen, and L. A. Tuan, "A Review on Implementation of Vehicle to Everything (V2X): Benefits, Barriers and Measures," in *2022 IEEE PES Innovative Smart Grid Technologies Conference Europe (ISGT-Europe)*, Novi Sad, Serbia, 2022, pp. 1–6, doi: 10.1109/ISGT-Europe54678.2022.9960673.

[27] S. Gyawali, S. Xu, Y. Qian, and R. Q. Hu, "Challenges and Solutions for Cellular Based V2X Communications," *IEEE Communications Surveys & Tutorials*, vol. 23, no. 1, pp. 222–255, 2021, doi: 10.1109/COMST.2020.3029723.

[28] J. Wang, Y. Shao, Y. Ge, and R. Yu, "A Survey of Vehicle to Everything (V2X) Testing," *Sensors*, early access. doi: 10.3390/s19020334.

[29] I. Soto, M. Calderon, O. Amador, and M. Urueña, "A survey on road safety and traffic efficiency vehicular applications based on C-V2X technologies," *Vehicular Communications*, vol. 33, p. 100428, 2022, doi: 10.1016/j.vehcom.2021.100428.

[30] A. Nair and S. Tanwar, "Resource allocation in V2X communication: State-of-the-art and research challenges," *PHYSICAL COMMUNICATION*, vol. 64, p. 102351, 2024, doi: 10.1016/j.phycom.2024.102351.

[31] I. Brahmi, H. Koubaa, and F. Zarai, "Resource allocation for Vehicle-to-Everything communications: A survey," *IET Networks*, vol. 12, no. 3, pp. 98–121, 2023, doi: 10.1049/ntw2.12078.

[32] A. Masmoudi, K. Mnif, and F. Zarai, "A Survey on Radio Resource Allocation for V2X Communication," *Wireless Communications and Mobile Computing*, vol. 2019, pp. 1–12, 2019, doi: 10.1155/2019/2430656.

[33] IEEE Author Center Journals. "Abstracting & Indexing (A&I) Databases - IEEE Author Center Journals." Accessed: Feb. 8, 2024.

[34] M. Templier and G. Paré, "A Framework for Guiding and Evaluating Literature Reviews," *CAIS*, vol. 37, 2015, doi: 10.17705/1CAIS.03706.

[35] H. Snyder, "Literature review as a research methodology: An overview and guidelines," *Journal of Business Research*, vol. 104, pp. 333–339, 2019, doi: 10.1016/j.jbusres.2019.07.039.

[36] R. Perez, F. Schubert, R. Rasshofer, and E. Biebl, "Deep Learning Radar Object Detection and Classification for Urban Automotive Scenarios," in *2019 Kleinheubach Conference*, 2019.

[37] E. E. González, D. Garcia-Roger, and J. F. Monserrat, "LTE/NR V2X Communication Modes and Future Requirements of Intelligent Transportation Systems Based on MR-DC Architectures," *Sustainability*, vol. 14, no. 7, p. 3879, 2022. doi: 10.3390/su14073879. [Online]. Available: https://www.mdpi.com/2071-1050/14/7/3879

[38] W. Anwar, N. Franchi, and G. Fettweis, "Physical Layer Evaluation of V2X Communications Technologies: 5G NR-V2X, LTE-V2X, IEEE 802.11bd, and IEEE 802.11p," in *2019 IEEE 90th Vehicular Technology Conference (VTC2019-Fall): Proceedings : Honolulu, Hawaii, USA 22-25 September 2019*, Honolulu, HI, USA, 2019, pp. 1–7, doi: 10.1109/VTCFall.2019.8891313.

[39] G. Naik, B. Choudhury, and J.-M. Park, "IEEE 802.11bd & 5G NR V2X: Evolution of Radio Access Technologies for V2X Communications," *IEEE Access*, vol. 7, pp. 70169–70184, 2019, doi: 10.1109/ACCESS.2019.2919489.

[40] F. Arena, G. Pau, and A. Severino, "A Review on IEEE 802.11p for Intelligent Transportation Systems," *JSAN*, vol. 9, no. 2, p. 22, 2020, doi: 10.3390/jsan9020022.

[41] B. Y. Yacheur, T. Ahmed, and M. Mosbah, "Analysis and Comparison of IEEE 802.11p and IEEE 802.11bd," in *Communication Technologies for Vehicles: 15th International Workshop, Nets4Cars/Nets4Trains/Nets4Aircraft 2020, Bordeaux, France, November 16–17, 2020, Proceedings* (Springer eBook Collection 12574), F. Krief, H. Aniss, L. Mendiboure, S. Chaumette, and M. Berbineau, Eds., 1st ed. Cham: Springer International Publishing; Imprint Springer, 2020, pp. 55–65.

[42] V. Mannoni, V. Berg, S. Sesia, and E. Perraud, "A Comparison of the V2X Communication Systems: ITS-G5 and C-V2X," in *2019 IEEE 89th Vehicular Technology Conference (VTC Spring): Proceedings : Kuala Lumpur, Malaysia, 28 April-1 May 2019*, Kuala Lumpur, Malaysia, 2019, pp. 1–5, doi: 10.1109/VTCSpring.2019.8746562.

[43] T. Shimizu, H. Lu, J. Kenney, and S. Nakamura, "Comparison of DSRC and LTE-V2X PC5 Mode 4 Performance in High Vehicle Density Scenarios," in *26th ITS World Congress, Singapore*, 2019. [Online]. Available: https://www.researchgate.net/publication/336768425_Comparison_of_DSRC_and_LTE-V2X_PC5_Mode_4_Performance_in_High_Vehicle_Density_Scenarios

[44] K. Ansari, "Joint use of DSRC and C-V2X for V2X communications in the 5.9 GHz ITS band," *IET intell. transp. syst*, vol. 15, no. 2, pp. 213–224, 2021. doi: 10.1049/itr2.12015. [Online]. Available: https://ietresearch.onlinelibrary.wiley.com/doi/pdf/10.1049/itr2.12015

[45] *ETSI EN 302 571: Intelligent Transport Systems (ITS); Radiocommunications equipment operating in the 5 855 MHz to 5 925 MHz frequency band; Harmonised Standard covering the essential requirements of article 3.2 of Directive 2014/53/EU,* ETSI EN 302 571, Feb. 2017. [Online]. Available: https://www.etsi.org/deliver/etsi_en/302500_302599/302571/02.01.01_60/en_302571v020101p.pdf

[46] K. Kiela *et al.,* "Review of V2X–IoT Standards and Frameworks for ITS Applications," *Applied Sciences*, vol. 10, no. 12, p. 4314, 2020, doi: 10.3390/app10124314.

[47] "Next Generation V2X – IEEE 802.11bd as fully backward compatible evolution of IEEE 802.11p," CAR 2 CAR Communication Consortium, Feb. 2023. Accessed: Jul. 15, 2024. [Online]. Available: https://www.car-2-car.org/fileadmin/documents/General_Documents/C2CCC_WP_2098_IEEE_802.11bd_TheV2XEvolution_V1.0.pdf

[48] V. Torgunakov, V. Loginov, and E. Khorov, "A Study of Channel Bonding in IEEE 802.11bd Networks," *IEEE Access*, vol. 10, pp. 25514–25533, 2022, doi: 10.1109/ACCESS.2022.3155814.

[49] R. Sattiraju, D. Wang, A. Weinand, and H. D. Schotten, "Link Level Performance Comparison of C-V2X and ITS-G5 for Vehicular Channel Models," in *2020 IEEE 91st Vehicular Technology Conference (VTC Spring): Proceedings : Antwerp, Belgium, 25-28 May 2020*, Antwerp, Belgium, 2020, pp. 1–7, doi: 10.1109/VTC2020-Spring48590.2020.9129366.

[50] "C-ITS FAQs: When will drivers benefit from cooperative V2X on European roads?" Accessed: May 24, 2022. [Online]. Available: https://www.car-2-car.org/about-c-its/c-its-faqs/

[51] M. Shi, Y. Zhang, D. Yao, and C. Lu, "Application-oriented performance comparison of 802.11p andLTE-V in a V2V communication system," *Tinshhua Sci. Technol.*, vol. 24, no. 2, pp. 123–133, 2019, doi: 10.26599/TST.2018.9010075.

[52] B. Fernandes, J. Rufino, M. Alam, and J. Ferreira, "Implementation and Analysis of IEEE and ETSI Security Standards for Vehicular Communications," *Mobile Netw Appl*, vol. 23, no. 3, pp. 469–478, 2018, doi: 10.1007/s11036-018-1019-x.

[53] M. N. Tahir and M. Katz, "Heterogeneous (ITS-G5 and 5G) Vehicular Pilot Road Weather Service Platform in a Realistic Operational Environment," *Sensors*, early access. doi: 10.3390/s21051676.

[54] M. Harounabadi, D. M. Soleymani, S. Bhadauria, M. Leyh, and E. Roth-Mandutz, "V2X in 3GPP Standardization: NR Sidelink in Release-16 and Beyond," *IEEE Comm. Stand. Mag.*, vol. 5, no. 1, pp. 12–21, 2021, doi: 10.1109/MCOMSTD.001.2000070.

[55] K. Ganesan, P. B. Mallick, J. Lohr, D. Karampatsis, and A. Kunz, "5G V2X Architecture and Radio Aspects," in *2019 IEEE Conference on Standards for Communications and Networking (CSCN): 2019 IEEE Conference on Standards for Communications and Networking (CSCN) took place 28-30 October 2019 in Granada, Spain*, Granada, Spain, 2019, pp. 1–6, doi: 10.1109/CSCN.2019.8931319.

[56] K. Sehla, T. M. T. Nguyen, G. Pujolle, and P. B. Velloso, "Resource Allocation Modes in C-V2X: From LTE-V2X to 5G-V2X," *IEEE Internet Things J.*, p. 1, 2022, doi: 10.1109/JIOT.2022.3159591.

[57] Z. Ali, S. Lagen, L. Giupponi, and R. Rouil, "3GPP NR V2X Mode 2: Overview, Models and System-Level Evaluation," *IEEE Access*, vol. 9, pp. 1–26, 2021, doi: 10.1109/ACCESS.2021.3090855.

[58] K. Jellid and T. Mazri, "DSRC vs LTE V2X for Autonomous Vehicle Connectivity," in *Innovations in Smart Cities Applications Volume 4: The Proceedings of the 5th International Conference on Smart City Applications* (Springer eBook Collection 183), M. Ben Ahmed, İ. Rakıp Karaş, D. Santos, O. Sergeyeva, and A. A. Boudhir, Eds., 1st ed. Cham: Springer International Publishing; Imprint Springer, 2021, pp. 381–394.

[59] R. Molina-Masegosa and J. Gozalvez, "LTE-V for Sidelink 5G V2X Vehicular Communications: A New 5G Technology for Short-Range Vehicle-to-Everything Communications," *IEEE Vehicular Technology Magazine*, vol. 12, no. 4, pp. 30–39, 2017, doi: 10.1109/MVT.2017.2752798.

[60] M. M. Saad, M. A. Tariq, J. Seo, and D. Kim, "An Overview of 3GPP Release 17 & 18 Advancements in the Context of V2X Technology," in *2023 International Conference on Artificial Intelligence in Information and Communication (ICAIIC)*, Bali, Indonesia, 2023, pp. 57–62, doi: 10.1109/ICAIIC57133.2023.10067121.

[61] X. Lin, "An Overview of 5G Advanced Evolution in 3GPP Release 18," *IEEE Comm. Stand. Mag.*, vol. 6, no. 3, pp. 77–83, 2022, doi: 10.1109/MCOMSTD.0001.2200001.

[62] M. Wang *et al.,* "Comparison of LTE and DSRC-Based Connectivity for Intelligent Transportation Systems," in *2017 IEEE 85th Vehicular Technology Conference (VTC Spring)*, 2017, pp. 1–5, doi: 10.1109/VTCSpring.2017.8108284.

[63] 3GPP. "Release 17." Accessed: Dec. 22, 2023. [Online]. Available: https://www.3gpp.org/specifications-technologies/releases/release-17

[64] 3GPP. "Release 18." Accessed: Dec. 22, 2023. [Online]. Available: https://www.3gpp.org/specifications-technologies/releases/release-18

[65] W. Lei *et al.,* "NR V2X Sidelink Design," in *5G System Design: An End to End Perspective* (Wireless Networks), W. Lei et al., Eds., Cham: Springer International Publishing, 2021, pp. 413–514.

[66] M. A. Khan *et al.,* "Level-5 Autonomous Driving—Are We There Yet? A Review of Research Literature," *ACM Comput. Surv.*, vol. 55, no. 2, pp. 1–38, 2022, doi: 10.1145/3485767.

[67] T. T. Thanh Le and S. Moh, "Comprehensive Survey of Radio Resource Allocation Schemes for 5G V2X Communications," *IEEE Access*, vol. 9, pp. 123117–123133, 2021, doi: 10.1109/ACCESS.2021.3109894.

[68] E. Cinque, F. Valentini, A. Persia, S. Chiocchio, F. Santucci, and M. Pratesi, "V2X Communication Technologies and Service Requirements for Connected and Autonomous Driving," in *2020 AEIT International Conference of Electrical and Electronic Technologies for Automotive (AEIT AUTOMOTIVE)*, Turin, Italy, uuuu-uuuu, pp. 1–6, doi: 10.23919/AEITAUTOMOTIVE50086.2020.9307388.

[69] A. Guizar, V. Mannoni, F. Poli, B. Denis, and V. Berg, "LTE-V2X performance evaluation for cooperative collision avoidance (CoCA) systems," in *2021 IEEE 93rd Vehicular Technology Conference (VTC2021-Spring)*, 2021, pp. 1–5, doi: 10.1109/VTC2020-Fall49728.2020.9348621.

[70] Andrea Ulbel, "Analysis of V2X Performance and Rollout Status with a Special Focus on Austria," Masterarbeit, Institute of Automation and Control, Graz Univerity of Technology, Graz, 2021. [Online]. Available: https://www.researchgate.net/publication/351443279_Analysis_of_V2X_Performance_and_Rollout_Status_with_a_Special_Focus_on_Austria

[71] R. Molina-Masegosa, J. Gozalvez, and M. Sepulcre, "Comparison of IEEE 802.11p and LTE-V2X: An Evaluation With Periodic and Aperiodic Messages of Constant and Variable Size," *IEEE Access*, vol. 8, pp. 121526–121548, 2020, doi: 10.1109/ACCESS.2020.3007115.

[72] V. Maglogiannis, D. Naudts, S. Hadiwardoyo, D. van den Akker, J. Marquez-Barja, and I. Moerman, "Experimental V2X Evaluation for C-V2X and ITS-G5 Technologies in a Real-Life Highway Environment," *IEEE Trans. Netw. Serv. Manage.*, p. 1, 2021, doi: 10.1109/TNSM.2021.3129348.

[73] M. Karoui, A. Freitas, and G. Chalhoub, "Performance comparison between LTE-V2X and ITS-G5 under realistic urban scenarios," in *2020 IEEE 91st Vehicular Technology Conference (VTC Spring): Proceedings : Antwerp, Belgium, 25-28 May 2020*, Antwerp, Belgium, 2020, pp. 1–7, doi: 10.1109/VTC2020-Spring48590.2020.9129423.

[74] G. Cecchini, A. Bazzi, B. M. Masini, and A. Zanella, "Performance comparison between IEEE 802.11p and LTE-V2V in-coverage and out-of-coverage for cooperative awareness," in *2017 IEEE Vehicular Networking Conference (VNC): 27-29 Nov. 2017*, Torino, O. Altintas, C. Casetti, N. Kirsch, R. Lo Cigno, and R. Meireles, Eds., 2017, pp. 109–114, doi: 10.1109/VNC.2017.8275637.

[75] A. Bazzi, G. Cecchini, M. Menarini, B. M. Masini, and A. Zanella, "Survey and Perspectives of Vehicular Wi-Fi versus Sidelink Cellular-V2X in the 5G Era," *Future Internet*, vol. 11, no. 6, p. 122, 2019, doi: 10.3390/fi11060122.

[76] V. Vukadinovic *et al.,* "3GPP C-V2X and IEEE 802.11p for Vehicle-to-Vehicle communications in highway platooning scenarios," *Ad Hoc Networks*, vol. 74, pp. 17–29, 2018. doi: 10.1016/j.adhoc.2018.03.004. [Online]. Available: https://www.sciencedirect.com/science/article/pii/S157087051830057X

[77] A. Bazzi, B. M. Masini, A. Zanella, and I. Thibault, "On the Performance of IEEE 802.11p and LTE-V2V for the Cooperative Awareness of Connected Vehicles," *IEEE Transactions on Vehicular Technology*, vol. 66, no. 11, pp. 10419–10432, 2017, doi: 10.1109/TVT.2017.2750803.

[78] P. Roux, S. Sesia, V. Mannoni, and E. Perraud, "System Level Analysis for ITS-G5 and LTE-V2X Performance Comparison," in *2019 IEEE 16th International Conference on Mobile Ad Hoc and Sensor Systems (MASS)*, Monterey, CA, USA, 2019, pp. 1–9, doi: 10.1109/MASS.2019.00010.

[79] T. Zugno, M. Drago, M. Giordani, M. Polese, and M. Zorzi, "NR V2X Communications at Millimeter Waves: An End-to-End Performance Evaluation," in *GLOBECOM 2020 - 2020 IEEE Global Communications Conference*, Taipei, Taiwan, uuuu-uuuu, pp. 1–6, doi: 10.1109/GLOBECOM42002.2020.9348259.

[80] M. M. Saad, M. T. R. Khan, S. H. A. Shah, and D. Kim, "Advancements in Vehicular Communication Technologies: C-V2X and NR-V2X Comparison," *IEEE Commun. Mag.*, vol. 59, no. 8, pp. 107–113, 2021, doi: 10.1109/MCOM.101.2100119.

[81] A. Rehman, R. Valentini, E. Cinque, P. Di Marco, and F. Santucci, "On the Impact of Multiple Access Interference in LTE-V2X and NR-V2X Sidelink Communications," *Sensors*, early access. doi: 10.3390/s23104901.

[82] P. Roux and V. Mannoni, "Performance evaluation for co-channel coexistence between ITS-G5 and LTE-V2X," in *2021 IEEE 93rd Vehicular Technology Conference (VTC2021-Spring)*, 2021, pp. 1–5, doi: 10.1109/VTC2020-Fall49728.2020.9348517.

[83] F. Abbas and P. Fan, "A Hybrid Low-Latency D2D Resource Allocation Scheme Based on Cellular V2X Networks," in *2018 IEEE International Conference on Communications Workshops (ICC Workshops)*, Kansas City, MO, USA, 2018, pp. 1–6, doi: 10.1109/ICCW.2018.8403512.

[84] C. A. Grazia, "On the Performance of IEEE 802.11p Outside the Context of a BSS Networks," in *2018 IEEE 29th Annual International Symposium on Personal, Indoor and Mobile Radio Communications (PIMRC)*, Bologna, Italy, 2018, pp. 1388–1393, doi: 10.1109/PIMRC.2018.8580785.

[85] S. H. Lim, Y. K. Chia, and L. Wynter, "Accurate and cost-effective traffic information acquisition using adaptive sampling: Centralized and V2V schemes," *Transportation Research Part C: Emerging Technologies*, vol. 94, pp. 99–120, 2018, doi: 10.1016/j.trc.2017.10.017.

[86] K. Thomas, H. Fouchal, S. Cormier, and F. Rousseaux, "C-ITS Communications based on BLE Messages," in *GLOBECOM 2020 - 2020 IEEE Global Communications Conference*, Taipei, Taiwan, uuuu-uuuu, pp. 1–7, doi: 10.1109/GLOBECOM42002.2020.9322076.

[87] M. Sepulcre, J. Gozalvez, G. Thandavarayan, B. Coll-Perales, J. Schindler, and M. Rondinone, "On the Potential of V2X Message Compression for Vehicular Networks," *IEEE Access*, vol. 8, pp. 214254–214268, 2020, doi: 10.1109/ACCESS.2020.3041688.

[88] C. Bin Ali Wael, N. Armi, A. Mitayani, D. Kurniawan, A. Suryadi Satyawan, and A. Subekti, "Analysis of IEEE 802.11p MAC Protocol for Safety Message Broadcast in V2V Communication," in *"Fostering innovation through ICTs for sustainable smart society": Proceeding : 2020 International Conference on Radar, Antenna, Microwave, Electronics and Telecommunications : virtual conference, 18-20 November 2020*, Tangerang, Indonesia, 2020, pp. 320–324, doi: 10.1109/ICRAMET51080.2020.9298654.

[89] R. Jacob, W. Anwar, N. Schwarzenberg, N. Franchi, and G. Fettweis, "System-level Performance Comparison of IEEE 802.11p and 802.11bd Draft in Highway Scenarios," in *2020 27th International Conference on Telecommunications (ICT)*, Bali, Indonesia, 2020, pp. 1–6, doi: 10.1109/ICT49546.2020.9239538.

[90] J. Hu et al., "Link level performance comparison between LTE V2X and DSRC," *J. Commun. Inf. Netw.*, vol. 2, no. 2, pp. 101–112, 2017. doi: 10.1007/s41650-017-0022-x. [Online]. Available: https://link.springer.com/article/10.1007/s41650-017-0022-x

[91] H. Xing, B. Cimoli, I. Passchier, G. Kakes, V. Ho, and H. Nijmeijer, "Practical Challenges in CACC Communication: ITS G5, LTE Uu, and LTE Sidelink PC5," in *2021 European Control Conference (ECC)*, Delft, Netherlands, 2021, pp. 1795–1801, doi: 10.23919/ECC54610.2021.9655171.

[92] J. Zhao, X. Gai, and X. Luo, "Performance Comparison of Vehicle Networking Based on DSRC and LTE Technology," in *2021 6th International Conference on Intelligent Transportation Engineering (ICITE 2021)* (Lecture Notes in Electrical Engineering), Z. Zhang, Ed., Singapore: Springer Nature Singapore, 2022, pp. 730–746.

[93] W. Anwar, S. Dev, A. Kumar, N. Franchi, and G. Fettweis, "PHY Abstraction Techniques for V2X Enabling Technologies: Modeling and Analysis," *IEEE Trans. Veh. Technol.*, vol. 70, no. 2, pp. 1501–1517, 2021, doi: 10.1109/TVT.2021.3053425.

[94] M. Montaño, R. Cajo, and F. Novillo, "Resource Allocation in C-V2X and DSRC Technologies: Analysis and Simulation-based Evaluation for V2V Direct Vehicular Communication," in *Proceedings of the Int'l ACM Symposium on Design and Analysis of Intelligent Vehicular Networks and Applications*, Montreal Quebec Canada, R. Coutinho and M. Bani Younes, Eds., 2023, pp. 115–122, doi: 10.1145/3616392.3623415.

[95] A. V. Pestryakov, E. R. Khasianova, and S. I. Dinges, "Analysis of Cellular Vehicle-to-Everything Physical Layer Parameters," in *2021 Systems of Signals Generating and Processing in the Field of on Board Communications*, Moscow, Russia, 2021, pp. 1–4, doi: 10.1109/IEEECONF51389.2021.9416036.

[96] N. Bonjorn, F. Foukalas, and P. Pop, "Enhanced 5G V2X services using sidelink device-to-device communications," in *2018 17th Annual Mediterranean Ad Hoc Networking Workshop (Med-Hoc-Net)*, Capri, 2018, pp. 1–7, doi: 10.23919/MedHocNet.2018.8407085.

[97] H. Su, M.-S. Pan, and S.-W. Kao, "A Subchannel Collision Reduction Method for 3GPP C- V2X Mode 4 Network Based on Vehicular Moving Status," in *2022 23rd Asia-Pacific Network Operations and Management Symposium (APNOMS)*, Takamatsu, Japan, 2022, pp. 1–6, doi: 10.23919/APNOMS56106.2022.9919919.

[98] J. Yin and S.-H. Hwang, "Variable MCS method for LTE V2V Mode4," in *2021 International Conference on Information and Communication Technology Convergence (ICTC)*, Jeju Island, Korea, Republic of, 2021, pp. 1368–1370, doi: 10.1109/ICTC52510.2021.9620930.

[99] F. Eckermann, M. Kahlert, and C. Wietfeld, "Performance Analysis of C-V2X Mode 4 Communication Introducing an Open-Source C-V2X Simulator," in *2019 IEEE 90th Vehicular Technology Conference (VTC2019-Fall): Proceedings : Honolulu, Hawaii, USA 22-25 September 2019*, Honolulu, HI, USA, 2019, pp. 1–5, doi: 10.1109/VTCFall.2019.8891534.

[100] J. Kim, G. Noh, T. Kim, H. Chung, and I. Kim, "Link-Level Performance Evaluation of mmWave 5G NR Sidelink Communications," in *2021 International Conference on Information and Communication Technology Convergence (ICTC)*, Jeju Island, Korea, Republic of, 2021, pp. 1482–1485, doi: 10.1109/ICTC52510.2021.9621128.

[101] I. Serunin, A. Pudeev, J.-Y. Hwang, S.-W. Lee, and A. Maltsev, "An Overview and Performance Analysis of CQI Reporting in 5G NR Sidelink," in *2022 13th International Conference on Information and Communication Technology Convergence (ICTC)*, Jeju Island, Korea, Republic of, 2022, pp. 34–39, doi: 10.1109/ICTC55196.2022.9952605.

[102] C. Campolo, V. Todisco, A. Molinaro, A. Berthet, S. Bartoletti, and A. Bazzi, "Improving Resource Allocation for beyond 5G V2X Sidelink Connectivity," in *2021 55th Asilomar Conference on Signals, Systems, and Computers*, Pacific Grove, CA, USA, 2021, pp. 55–60, doi: 10.1109/IEEECONF53345.2021.9723155.

[103] A. V. Olteanu and M. Nicolae, "Using advanced V2X communication technologies in self-organized VANETs," in *2021 23rd International Conference on Control Systems and Computer Science (CSCS)*, 2021, pp. 254–259, doi: 10.1109/CSCS52396.2021.00049.

[104] United States Department of Transportation. "Intelligent Transportation Systems - Communications." Accessed: Jun. 21, 2022. [Online]. Available: https://www.its.dot.gov/press/2018/v2x.htm

[105] Y. Yang and K. Hua, "Emerging Technologies for 5G-Enabled Vehicular Networks," *IEEE Access*, vol. 7, pp. 181117–181141, 2019, doi: 10.1109/ACCESS.2019.2954466.

[106] L. Miao, S.-F. Chen, Y.-L. Hsu, and K.-L. Hua, "How Does C-V2X Help Autonomous Driving to Avoid Accidents?," *Sensors (Basel, Switzerland)*, early access. doi: 10.3390/s22020686.

[107] Federal Communications Commision, *FCC Modernizes 5.9 GHz Band For Wi-Fi And Auto Safety*, 2020. Accessed: Aug. 18, 2022. [Online]. Available: https://www.fcc.gov/document/fcc-modernizes-59-ghz-band-improve-wi-fi-and-automotive-safety

[108] Audi. "Audi Newsroom." Accessed: Jun. 21, 2022. [Online]. Available: https://media.audiusa.com/en-us/releases/494

[109] "Strategic Analysis of the Global Vehicle-to-Everything (V2X) Market, Forecast to 2025," Frost & Sullivan, Rep. ME50-18, Aug. 2020.

[110] 5G Americas, Ed., "Vehicular Connectivity: C-V2X & 5G: A 5G Americas White Paper," Sep. 2021. Accessed: Jun. 18, 2022. [Online]. Available: https://www.5gamericas.org/wp-content/uploads/2021/09/Vehicular-Connectivity-C-V2X-and-5G-InDesign-1.pdf

[111] Ofinno. "5G Cellular Technology for Connected Cars Receives Boost from European Legislators - Ofinno." Accessed: Jun. 23, 2022. [Online]. Available: https://ofinno.com/article/5g-cellular-technology-for-connected-cars-receives-boost-from-european-legislators/

[112] *Intelligent Transport Systems (ITS); LTE-V2X Access layer specification for Intelligent Transport Systems operating in the 5 GHz frequency band,* European Telecommunications Standards Institute. [Online]. Available: https://www.etsi.org/deliver/etsi_en/303600_303699/303613/01.01.01_30/en_303613v010101v.pdf

[113] C2C-CC, Ed., "ITS-G5 and Sidelink LTE-V2X Co-Channel Coexistence Mitigation Methods," Apr. 2021. Accessed: Jun. 18, 2022. [Online]. Available: https://www.car-2-car.org/fileadmin/documents/General_Documents/C2CCC_WP_2091_Co-ChannelCoexistence_MitigationMethods_V1.0.pdf

[114] M. Demler. "C-V2X drives intelligent transportation." [Online]. Available: https://www.linleygroup.com/uploads/qualcomm-c-v2x-technology-wp.pdf

[115] "Euro NCAP 2025 Roadmap: In pursuit of vision zero," Euro NCAP, 12.Sept.2017. Accessed: Oct. 20, 2023. [Online]. Available: https://cdn.euroncap.com/media/30700/euroncap-roadmap-2025-v4.pdf

[116] "Euro NCAP Vision 2030: A safer future for mobility," Euro NCAP, Nov. 2022. Accessed: Oct. 20, 2023. [Online]. Available: https://cdn.euroncap.com/media/74468/euro-ncap-roadmap-vision-2030.pdf

[117] L. Miao, J. J. Virtusio, and K.-L. Hua, "PC5-Based Cellular-V2X Evolution and Deployment," *Sensors (Basel, Switzerland)*, early access. doi: 10.3390/s21030843.

[118] ResearchAndMarkets. "China Automotive and 5G Industry Integration Development Report 2022: Application of 5G-V2X Will Promote the Realization of High-level Autonomous Driving - ResearchAndMarkets.com." Accessed: Jun. 21, 2022. [Online].

Available: https://www.businesswire.com/news/home/ 20220523005608/en/China-Automotive-and-5G-Industry-Integration-Development-Report-2022-Application-of-5G-V2X-Will-Promote-the-Realization-of-High-level-Autonomous-Driving---ResearchAndMarkets.com

[119] T. Cui, L. Li, Z. Zhang, and C. Sun, "C-V2X Vision in the Chinese Roadmap: Standardization, Field Tests, and Industrialization," in *Vehicular Networks - Principles, Enabling Technologies and Perspectives [Working Title]*, IntechOpen, 2022.

[120] G. Thandavarayan, M. Sepulcre, and J. Gozalvez, "Redundancy Mitigation in Cooperative Perception for Connected and Automated Vehicles," in *2021 IEEE 93rd Vehicular Technology Conference (VTC2021-Spring)*, 2021, pp. 1–5, doi: 10.1109/VTC2020-Spring48590.2020.9129445.

[121] G. Cui, W. Zhang, Y. Xiao, L. Yao, and Z. Fang, "Cooperative Perception Technology of Autonomous Driving in the Internet of Vehicles Environment: A Review," *Sensors*, early access. doi: 10.3390/s22155535.

[122] H. Huang, H. Li, C. Shao, T. Sun, W. Fang, and S. Dang, "Data Redundancy Mitigation in V2X Based Collective Perceptions," *IEEE Access*, vol. 8, pp. 13405–13418, 2020, doi: 10.1109/ACCESS.2020.2965552.

[123] Q. Delooz, R. Riebl, A. Festag, and A. Vinel, "Design and Performance of Congestion-Aware Collective Perception," in *2020 IEEE Vehicular Networking Conference (VNC)*, New York, NY, USA, 2020, pp. 1–8, doi: 10.1109/VNC51378.2020.9318335.

[124] *ETSI EN 302 637-2 - V1.4.1: Intelligent Transport Systems (ITS); Vehicular Communications; Basic Set of Applications; Part 2: Specification of Cooperative Awareness Basic Service,* European Telecommunications Standards Institute, Apr. 2019. [Online]. Available: https://www.etsi.org/deliver/etsi_en/302600_302699/ 30263702/01.04.01_60/en_30263702v010401p.pdf

[125] *ETSI EN 302 637-3 - V1.3.1: Intelligent Transport Systems (ITS); Vehicular Communications; Basic Set of Applications; Part 3: Specifications of Decentralized Environmental Notification Basic Service,* European Telecommunications Standards Institute, Apr. 2019. [Online]. Available: https://www.etsi.org/deliver/etsi_en/302600_ 302699/30263703/01.03.01_60/en_30263703v010301p.pdf

[126] *ETSI TR 103 562 - V2.1.1.: Intelligent Transport Systems (ITS); Vehicular Communications; Basic Set of Applications;Analysis of the Collective Perception Service (CPS); Release 2,* European Telecommunications Standards Institute, Dec. 2019. [Online]. Available: https://www.etsi.org/deliver/etsi_tr/103500_103599/ 103562/02.01.01_60/tr_103562v020101p.pdf

[127] *ETSI TS 103 301 - V1.3.1: Intelligent Transport Systems (ITS); Vehicular Communications; Basic Set of Applications; Facilities layer protocols and communication requirements for infrastructure services,* European Telecommunications Standards Institute, Feb. 2020. [Online]. Available: https://www.etsi.org/deliver/etsi_ts/103300_ 103399/103301/01.03.01_60/ts_103301v010301p.pdf

[128] X. Zheng, S. Li, Y. Li, D. Duan, L. Yang, and X. Cheng, "Confidence Evaluation for Machine Learning Schemes in Vehicular Sensor Networks," *IEEE Trans. Wireless Commun.*, vol. 22, no. 4, pp. 2833–2846, 2023, doi: 10.1109/TWC.2022.3214499.

[129] Q. Chen, S. Tang, Q. Yang, and S. Fu, "Cooper: Cooperative Perception for Connected Autonomous Vehicles based on 3D Point Clouds," May. 2019, doi: 39th. [Online]. Available: https://arxiv.org/ pdf/1905.05265

[130] Y. Endo and S. Kamijo, "Deep Voxelized Feature Maps for Self-Localization in Autonomous Driving," *Sensors*, early access. doi: 10.3390/s23125373.

[131] L. Ballotta, G. Peserico, F. Zanini, and P. Dini, "To Compute or not to Compute? Adaptive Smart Sensing in Resource-Constrained Edge Computing," 2022, doi: 10.48550/ARXIV.2209.02166.

[132] J. Wang, J. Liu, and N. Kato, "Networking and Communications in Autonomous Driving: A Survey," *IEEE Communications Surveys & Tutorials*, vol. 21, no. 2, pp. 1243–1274, 2019, doi: 10.1109/COMST.2018.2888904.

[133] F. Dettinger, H. Wei, M. Weiß, N. Jazdi, and M. Weyrich, "Dateneffiziente Vervollständigung des Umgebungsmodells von autonomen vernetzten Systemen mittels Sensorfusion," in *Automation 2023: Transformation by Automation, 24. Leitkongress der Mess- und Automatisierungstechnik, 27. und 28. Juni 2023, Baden-Baden* (VDI-Berichte 2419), Düsseldorf: VDI Verlag; VDI eLibrary, 2023, pp. 513–524.

[134] F. Hawlader and R. Frank, "Towards a Framework to Evaluate Cooperative Perception for Connected Vehicles," in *2021 IEEE Vehicular Networking Conference (VNC)*, Ulm, Germany, 2021, pp. 36–39, doi: 10.1109/VNC52810.2021.9644667.

[135] Q. Yang, S. Fu, H. Wang, and H. Fang, "Machine-Learning-Enabled Cooperative Perception for Connected Autonomous Vehicles: Challenges and Opportunities," *IEEE Network*, vol. 35, no. 3, pp. 96–101, 2021, doi: 10.1109/MNET.011.2000560.

[136] Q. Delooz *et al.*, "Analysis and Evaluation of Information Redundancy Mitigation for V2X Collective Perception," *IEEE Access*, vol. 10, pp. 47076–47093, 2022, doi: 10.1109/ACCESS.2022.3170029.

[137] G. Thandavarayan, M. Sepulcre, and J. Gozalvez, "Analysis of Message Generation Rules for Collective Perception in Connected and Automated Driving," in *IV19: 30th IEEE Intelligent Vehicles Symposium : 9-12 June 2019, Paris*, Paris, France, 2019, pp. 134–139, doi: 10.1109/IVS.2019.8813806.

[138] G. Thandavarayan, M. Sepulcre, and J. Gozalvez, "Generation of Cooperative Perception Messages for Connected and Automated Vehicles," *IEEE Trans. Veh. Technol.*, vol. 69, no. 12, pp. 16336–16341, 2020, doi: 10.1109/TVT.2020.3036165.

[139] A. Balador *et al.*, "Survey on decentralized congestion control methods for vehicular communication," *Vehicular Communications*, vol. 33, p. 100394, 2022, doi: 10.1016/j.vehcom.2021.100394.

[140] B. Kar, W. Yahya, Y.-D. Lin, and A. Ali, "Offloading Using Traditional Optimization and Machine Learning in Federated Cloud–Edge–Fog Systems: A Survey," *IEEE Communications Surveys & Tutorials*, vol. 25, no. 2, pp. 1199–1226, 2023, doi: 10.1109/COMST.2023.3239579.

[141] A, G. P. Wijesiri N. B., T. Samarasinghe, and J. Haapola, "Performance Enhancement of C-V2X Mode 4 Utilizing Multiple Candidate Single-subframe Resources," Feb. 2022. [Online]. Available: http://arxiv.org/pdf/2202.10869.pdf

[142] F. Abbas, X. Yuan, M. S. Bute, and P. Fan, "Performance Analysis Using Full Duplex Discovery Mechanism in 5G-V2X Communication Networks," *IEEE Trans. Intell. Transport. Syst.*, vol. 23, no. 8, pp. 11453–11464, 2022, doi: 10.1109/TITS.2021.3103974.

[143] F. Milani, M. Foell, and C. Beidl, "A Data-based Approach to Predict the Response Time of Cloud-based Vehicle Functions," in *2019 IEEE International Conference on Connected Vehicles and Expo (ICCVE)*, Graz, Austria, 2019, pp. 1–6, doi: 10.1109/ICCVE45908.2019.8965217.

[144] H. Wu, "Multi-Objective Decision-Making for Mobile Cloud Offloading: A Survey," *IEEE Access*, vol. 6, pp. 3962–3976, 2018, doi: 10.1109/ACCESS.2018.2791504.

[145] J. Liu, S. Wang, J. Wang, C. Liu, and Y. Yan, "A Task Oriented Computation Offloading Algorithm for Intelligent Vehicle Network With Mobile Edge Computing," *IEEE Access*, vol. 7, pp. 180491–180502, 2019, doi: 10.1109/ACCESS.2019.2958883.

[146] L. Gillam, K. Katsaros, M. Dianati, and A. Mouzakitis, "Exploring edges for connected and autonomous driving," in *IEEE INFOCOM 2018 - IEEE Conference on Computer Communications Workshops (INFOCOM WKSHPS)*, Honolulu, HI, 2018, pp. 148–153, doi: 10.1109/INFOCOMW.2018.8406890.

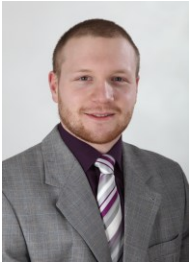
**Falk Dettinger, M.Sc.** is a PhD student and research assistant at the Institute of Industrial Automation and Software Engineering (IAS) at the University of Stuttgart, Germany. In addition, he is a lecturer in the Faculty of Electrical Engineering at the Baden-Wuerttemberg Cooperative State University (DHBW) Stuttgart, Germany.
He received his bachelor's degree in electrical engineering in 2017 from the DHBW Stuttgart, Germany and subsequently joined the Broetje Automation GmbH as commissioning engineer in the electrical engineering department. In 2019, he became a student at the University of Stuttgart, Germany, where he finished his master's degree with honors in 2021. His areas of interest include function offloading between connected vehicles and distributed cloud/edge systems and intelligent function orchestration.

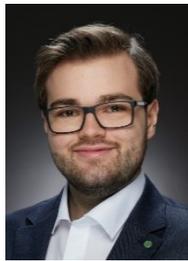
**Matthias Weiß, M.Sc.** is a PhD student and research assistant at the Institute of Industrial Automation and Software Engineering (IAS) at the University of Stuttgart, Germany. He received both his bachelor's and master's degree in electrical engineering from the University of Stuttgart in 2018 and 2021 respectively. His fields of research are the continuous analysis and optimization of distributed cloud/edge systems with the goal of reducing the operational complexity caused by frequent update deployments.

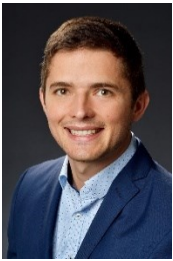
**Daniel Dittler, M.Sc.** is a PhD student and research assistant at the Institute of Industrial Automation and Software Engineering (IAS) at the University of Stuttgart, Germany. He received his bachelor's degree in mechanical engineering from Pforzheim University in 2019 and his master's degree in mechatronic system development in 2021. His areas of interest are digital twins in automation technology and the automatic model adaption of behavior models in the operational phase.

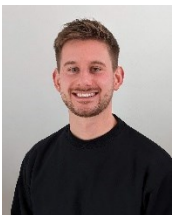
**Johannes Stümpfle, M.Sc.** is a PhD student and research assistant at the Institute of Industrial Automation and Software Engineering (IAS) at the University of Stuttgart, Germany. He received his bachelor's and master's degree in electrical engineering at the University of Stuttgart. His areas of interest include the evolution of software-intensive systems and mastering the complexity in highly variant automation systems.

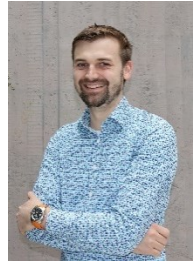
**Maurice Artelt, M.Sc.** is a PhD student and academic employee at the Institute of Industrial Automation and Software Engineering (IAS) at the University of Stuttgart, Germany. He received his bachelor's degree in electrical engineering in 2017 from the DHBW Mosbach, Germany and subsequently joined Bosch Rexroth AG, Erbach as industrie 4.0 production planer. In 2018, he additionally became a student at the University of Stuttgart, Germany, where he finished his master's degree in 2021.
His areas of interest include remaining useful life prediction and anomaly detection with hybrid prediction models for electrical function groups.

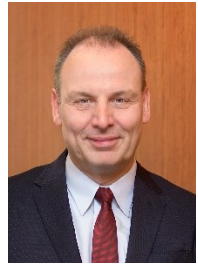
**Prof. Dr-Ing. Michael Weyrich** studied electrical engineering, specialising at the University of Applied Science Saarbrücken, University of Westminster (London, U.K.) and at the Ruhr-University Bochum, focussing on automation technology. Thereafter, he worked as a Research Assistant with Prof. Dr.-Ing. Paul Drews of the European Centre of Mechatronics and was awarded a doctoral degree in 1999 from the RWTH Aachen.
Prof. Weyrich worked with Daimler AG for eight years, where he assumed the leadership of the "End-to-End Process Flexible Manufacturing" project and later had a technical management function in the "CAx Process Chain – Production" of the Information Technology Management division. From 2004 onward, he was the head of "IT for Engineering" at DaimlerChrysler Research and Technology Bangalore (India). After his return to Germany, he joined Siemens AG as department leader with direct report to the Business Unit Head of Motion Control in Erlangen for 2 years, focussing on innovative technologies.
In 2009, he was appointed as Professor for the Chair of Automation in Manufacturing at the University of Siegen and as Director of the Automotive Centre Südwestfalen. Prof. Weyrich assumed the role of Director of the Institute of Industrial Automation and Software Engineering at the University of Stuttgart in 2013.
He is very interested in research in cyber physical systems for industrial application. He has published over 100 papers and is very active in research and is a member of the Board of the German Engineering foundations VDI/VDE GMA. Additionally, he is an appointed reviewer for the European Commission, the German Research Foundation, DFG and a number of other institutions.